\documentclass[aps,prl,twocolumn,amsmath,amssymb,floatfix,superscriptaddress,citeautoscript]{revtex4-2}
\usepackage{amssymb}
\usepackage{amsmath}
\usepackage{graphicx}
\usepackage{dcolumn}
\usepackage{hyperref}
\usepackage{siunitx}
\usepackage{bm}
\usepackage{xcolor}
\usepackage{natbib}

\def\etal.{et\penalty50\ al.}

\usepackage[commentmarkup=todo, authormarkup=none, highlightmarkup=background]{changes}
\definechangesauthor[name = Timothee, color=blue]{T}
\definechangesauthor[name = Georg, color=orange]{GK}

\newcommand{\pheliqs}{Univ. Grenoble Alpes, CEA, Grenoble INP, IRIG, PHELIQS, F-38000 Grenoble, France}
\newcommand{\neel}{Univ. Grenoble Alpes, CNRS, Institut Néel, F-38000 Grenoble, France}

\newcommand{\TU}{Institute for Solid State and Materials Physics, TU Dresden University of Technology, 01062 Dresden, Germany}
\newcommand{\MPG}{Max Planck Institute for Chemical Physics of Solids, 01187 Dresden, Germany}
\newcommand{\KIT}{Institute for Quantum Materials and Technologies, Karlsruhe Institute of Technology, Kaiserstra\ss e 12, 76131 Karlsruhe, Germany }

\begin{document}

\title{Quantitative thermodynamic study of superconducting and normal states  in \texorpdfstring{UTe$_2$}{UTe2} under pressure}
 
\author{T.~Vasina}
\thanks{These authors contributed equally to this work.}
\affiliation{\pheliqs}
\author{M.~Pfeiffer}
\thanks{These authors contributed equally to this work.}
\affiliation{\KIT}
\affiliation{\MPG}
\affiliation{\TU}
\author{R.~Borth}
\affiliation{\MPG}
\author{M.~Nicklas}
\affiliation{\MPG}
\author{M.~Amano Patino}
\affiliation{\pheliqs}
\affiliation{\neel}
\author{G.~Lapertot}
\affiliation{\pheliqs}
\author{J.-P.~Brison}
\affiliation{\pheliqs}
\author{E.~Hassinger}
\affiliation{\KIT}
\affiliation{\MPG}
\author{G.~Knebel}
\email[E-mail me at: ]{georg.knebel@cea.fr}
\affiliation{\pheliqs}
\author{D.~Braithwaite}
\email[E-mail me at: ]{daniel.braithwaite@cea.fr}
\affiliation{\pheliqs}

\begin{abstract}
We report a quantitative calorimetric study of UTe$_2$ under pressure with a direct measurement of the Sommerfeld $\gamma$ coefficient, showing a three-fold enhancement of electronic effective mass when approaching the critical pressure where superconductivity is suppressed and ordered states occur. We analyse the evolution of $\gamma$ with the amplitude of the jumps in the specific heat at the two superconducting transitions, and the superconducting critical temperature with pressure. This analysis would suggest that the high pressure superconducting phase nucleates only on a fraction of the Fermi surface. It also points to the possible major role of a quantum critical point of the  unidentified phase that has been called weak magnetic order, rather than to the critical pressure of the antiferromagnetic phase. Just at the border of long-range antiferromagnetic order, where superconductivity emerges from the weak magnetic order phase, a significant increase in the specific heat jump for both superconducting transitions is found, accompanied by a noticeable change of their shapes. 
\end{abstract}

\maketitle
In the landscape of unconventional superconductors, UTe$_2$ stands out as a remarkable material. Not only is it a very probable candidate for stabilizing spin-triplet superconductivity \cite{Ran2019,Aoki2019}, but it also exhibits several distinct superconducting states, a rare feature observed in only a handful of compounds \cite{Khim2021, Sauls1994, Ott1985}. These distinct superconducting phases can be induced by tuning the magnetic field \cite{Rosuel2023}, applying hydrostatic pressure \cite{Braithwaite2019, honda_pressure-induced_2023}, or varying the direction of the magnetic field \cite{Ran2019a}. Each superconducting phase of UTe$_2$ is thought to be stabilized by a distinct mechanism \cite{Rosuel2023, Wu2025}.

 When applying pressure on UTe$_2$, a new superconducting phase (SC2) is induced above $\SI{0.2}{\giga\pascal}$, and its critical temperature reaches a maximum near $\SI{1}{\giga\pascal}$ with a higher $T_\text{c}$ than that of the superconducting state SC1 at ambient pressure. When further increasing pressure to a critical pressure $p_\mathrm{c}$, superconductivity is abruptly suppressed by a first-order like transition, and a long-range antiferromagnetic (AFM) order is promoted below $T_N \approx 3.5$~K \cite{knafo_incommensurate_2025}. Another phase transition is detected by resistivity \cite{Ran2020, Aoki2021}, magnetization \cite{Li2021}, and calorimetry \cite{Thomas2020} at a higher temperature. This phase was proposed to be a short-range magnetic order and named “Weak Magnetic Order” (WMO), although its microscopic nature is yet to be determined. It was  recently shown that the pressure-induced phase SC2 is identical to the high-field-induced superconducting phase for $\bm{H}$ along the $\bm{b}$-axis \cite{vasina_connecting_2025}. This field-induced phase for $\bm{H}\parallel\bm{b}$ has been extensively studied, with a  quantitative specific heat study bringing  important information on the superconducting order parameter and pairing mechanism \cite{Rosuel2023}. In addition to clearly showing the existence of two distinct superconducting phases, this study revealed an increase of the normal state Sommerfeld coefficient $\gamma$ by a factor of two with increasing field up to the metamagnetic transition at $H_m \approx 34.5$~T. This indicates an enhanced effective mass $m^*$ from magnetic fluctuations on approaching $H_m$ \cite{Tokunaga2023}. The same mechanism controlling this mass enhancement might also trigger the increase of the superconducting transition with field for $\bm{H} \parallel \bm{b}$. This also suggests a different pairing mechanism of SC2 from that responsible for low field superconductivity SC1. This quantitative study allowed a careful analysis of the height and width of the jumps in $C/T$ related to the two superconducting transitions, providing insights into the SC2 superconducting phase. 
As SC2 can also be stabilized by pressure, a quantitative calorimetric study under pressure could similarly help refine our understanding of this SC2 phase. 
 
 Several ac calorimetry studies under pressure have already been performed \cite{Braithwaite2019, Thomas2020, Aoki2020, Wu2025PRL}. They revealed the existence of two superconducting phases, the appearance of the AFM and WMO phases, and some qualitative information on changes in the size of the jumps of $C/T$ at the onset of superconductivity. However ac calorimetry under pressure does not provide reliable quantitative values of the specific heat. A qualitative indication of enhanced electronic correlations on increasing pressure is seen in transport properties \cite{Valiska2021, Braithwaite2019, Thomas2020}, as well as microscopic nuclear magnetic resonance (NMR) measurements \cite{Kinjo2023PRB, Ambika2026}. A reliable determination of an increase of $\gamma$ under pressure, revealing the enhancement of the density of states and effective mass $m^*$ of the quasi-particles, together with its relationship to the appearance of AFM and WMO orders, and possible quantum critical effects, is lacking. A quantitative determination of the evolution of the relative $C/T$ jumps at the superconducting transitions would also clarify the superconducting phase diagram, namely the fact that three second-order transition lines meet in one point, and give more insight on the SC2 phase. 

\begin{figure}[t]
\centering
\includegraphics[width=\linewidth]{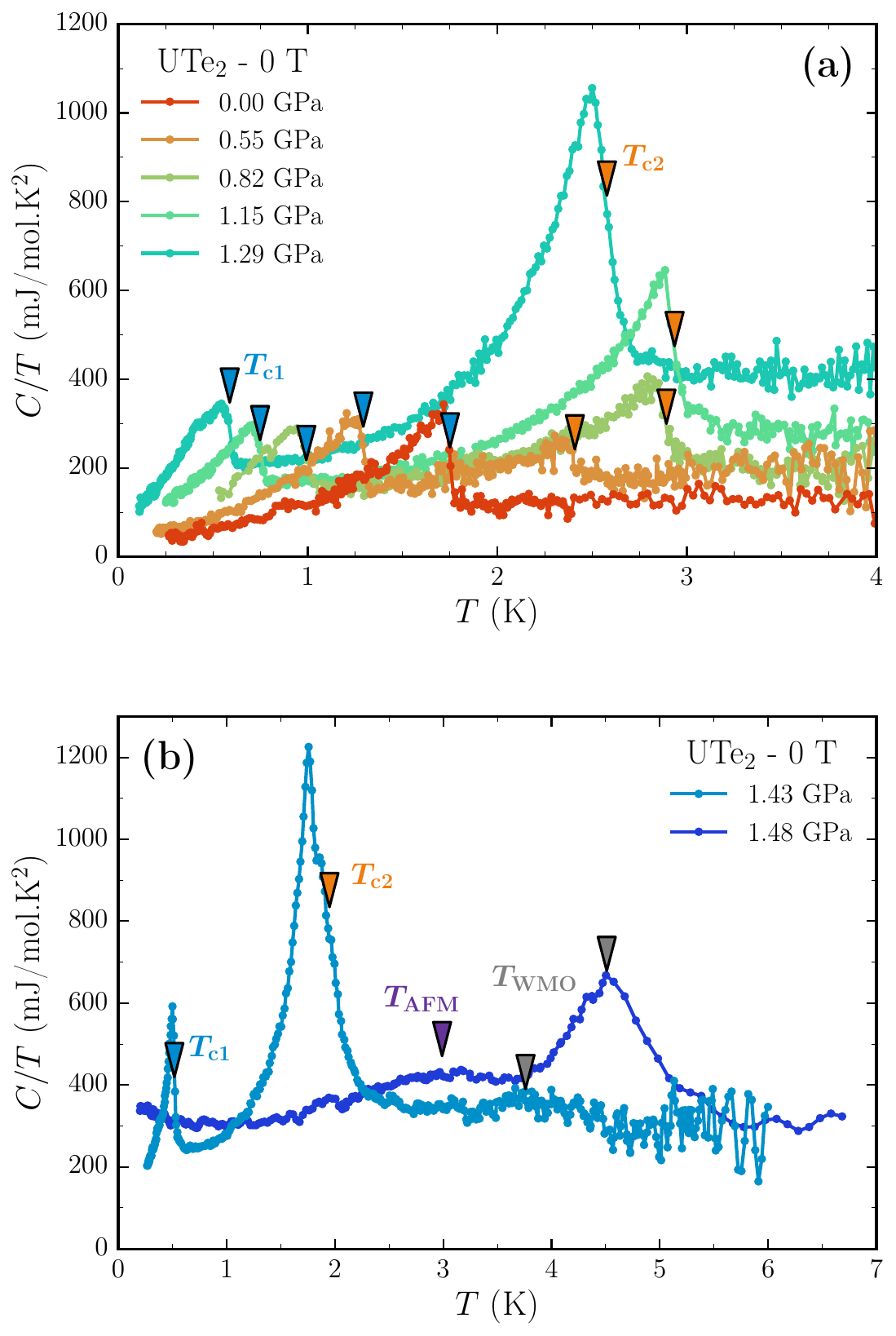}
\caption{\label{fig:raw}Specific heat as a function of temperature at different pressures. Blue and orange triangles represent superconducting transitions (to resp. SC1 and SC2). Violet and grey triangles represent the magnetic phases (resp. AFM and WMO).}
\end{figure}

Here, we present a quantitative specific heat study of UTe$_2$ under pressure. In order to preserve a sizable sample contribution against the addenda, we used a rather large single crystal ($m=\SI{23.73}{\milli\gram}$), hence grown by Chemical Vapor Transport (CVT).  It was placed in a small piston-cylinder pressure cell. The total heat capacity of the sample and pressure cell was measured using a quasi-adiabatic technique \cite{Nicklas2015}. The specific heat of UTe$_2$ was obtained by subtracting the contribution of the pressure cell, pressure transmitting medium (Fluorinert FC-72), and teflon cap, taking into account pressure effects on the specific heat of the latter two. Before the experiment, the specific heat of the sample was measured at ambient pressure. This measurement, and more details on the methods, are given in the Supplemental Material \cite{suppl}. Measurements were performed from close to ambient pressure up to $\SI{1.48}{\giga\pascal}$, above the critical pressure $p_\text{c}$ where long-range AFM order occurs.

The results of this study are shown in Fig.~\ref{fig:raw}. At nearly ambient pressure, a single phase transition is observed at about $T_\text{c1} = \SI{1.75}{\kelvin}$, corresponding to the transition between the paramagnetic phase and the superconducting SC1 phase. When pressure is increased to $\SI{0.55}{\giga\pascal}$, a second transition appears at higher temperatures, corresponding to the pressure-induced SC2 phase. At first, $T_\text{c2}$ is enhanced by pressure, but above $\approx \SI{1}{\giga\pascal}$, $T_\text{c2}$ starts decreasing. When reaching $p=\SI{1.43}{\giga\pascal}$, $T_\text{c1}$ and $T_\text{c2}$ are at their lowest. At this pressure, a broad anomaly has appeared at higher temperatures, corresponding to the WMO phase. Finally, at the highest pressure of $\SI{1.48}{\giga\pascal}$, the sample is above the critical pressure and superconductivity is completely suppressed, while antiferromagnetic order is stabilized at $T_\text{AFM} = \SI{3.3}{\kelvin}$. The anomaly associated with the magnetic WMO phase is now characterized by a larger specific heat jump, at around $\SI{4.5}{\kelvin}$. 

\begin{figure}
    \centering
    \includegraphics[width=0.85\linewidth]{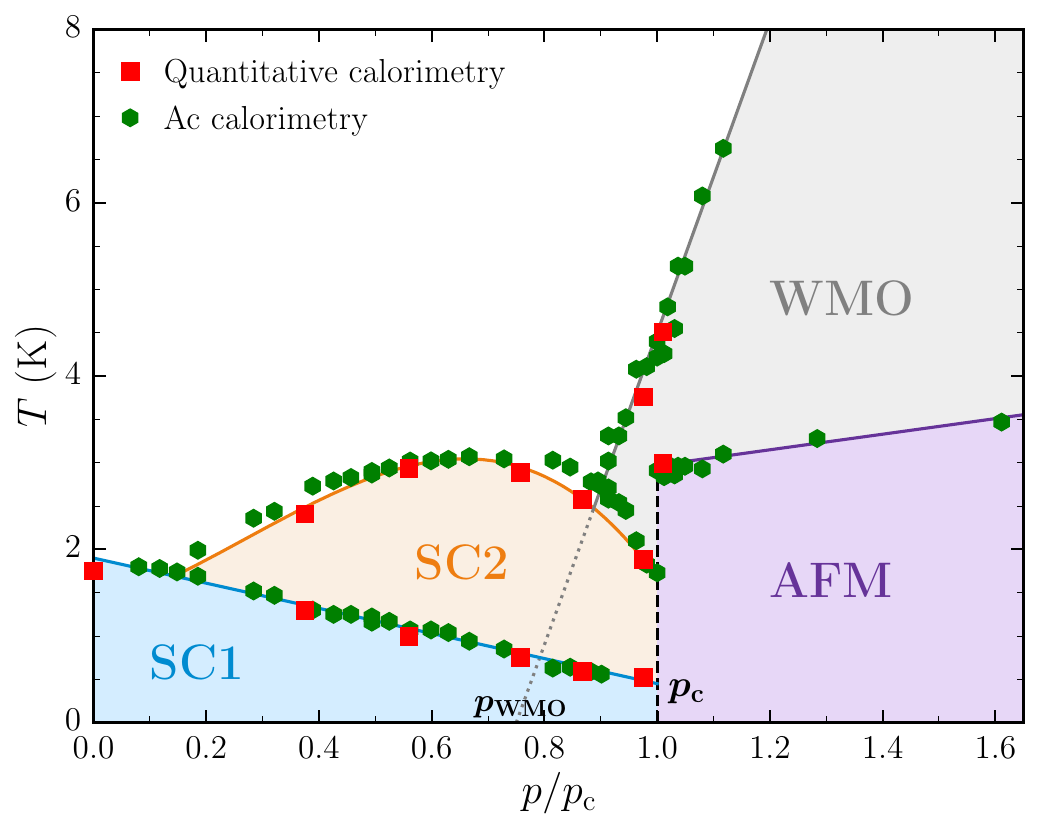}
    \caption{Zero-field $p-T$ phase diagram of UTe$_2$ obtained from  quantitative specific heat measurements in the piston cylinder cell (red squares), and ac calorimetry in the diamond anvil cell (green circles). The pressure is normalized to the critical pressure where the AFM order appears (1.45 GPa and 1.62 GPa for the piston-cylinder and DAC respectively. The WMO phase is stabilized in a significant pressure range before the critical pressure where the AFM state appears, with the maximum $T_\text{c}$ lying close to the extrapolated critical pressure $p_\text{WMO}$.}
    \label{fig:P-T}
\end{figure}

 These results indicate that the WMO phase is stabilized above the superconducting SC2 phase, at pressures clearly below the antiferromagnetic critical pressure. We also confirmed this on a high-quality MSF sample measured in a diamond anvil cell with in-situ pressure tuning to probe the phase diagram in more details. The resulting phase diagram for the two studies  is presented in Fig.~\ref{fig:P-T}. Because of the different experimental conditions, the pressure scale is  normalized to the critical pressure $p_{\rm c}$ of the AFM long range order for each case. This shows a significant pressure range where both, superconductivity and the WMO phase are found. It suggests that the WMO transition temperatures can be extrapolated down to $\SI{0}{\kelvin}$, at a putative critical pressure $p_\text{WMO}$, here equal to roughly $\SI{1.2}{\giga\pascal}$, close to the maximum of SC2. However no sign of the WMO transition was observed within the superconducting domain, suggesting that the onset of superconductivity could expel the WMO phase.

\begin{figure}[t]
\centering
\includegraphics[width=0.95\linewidth]{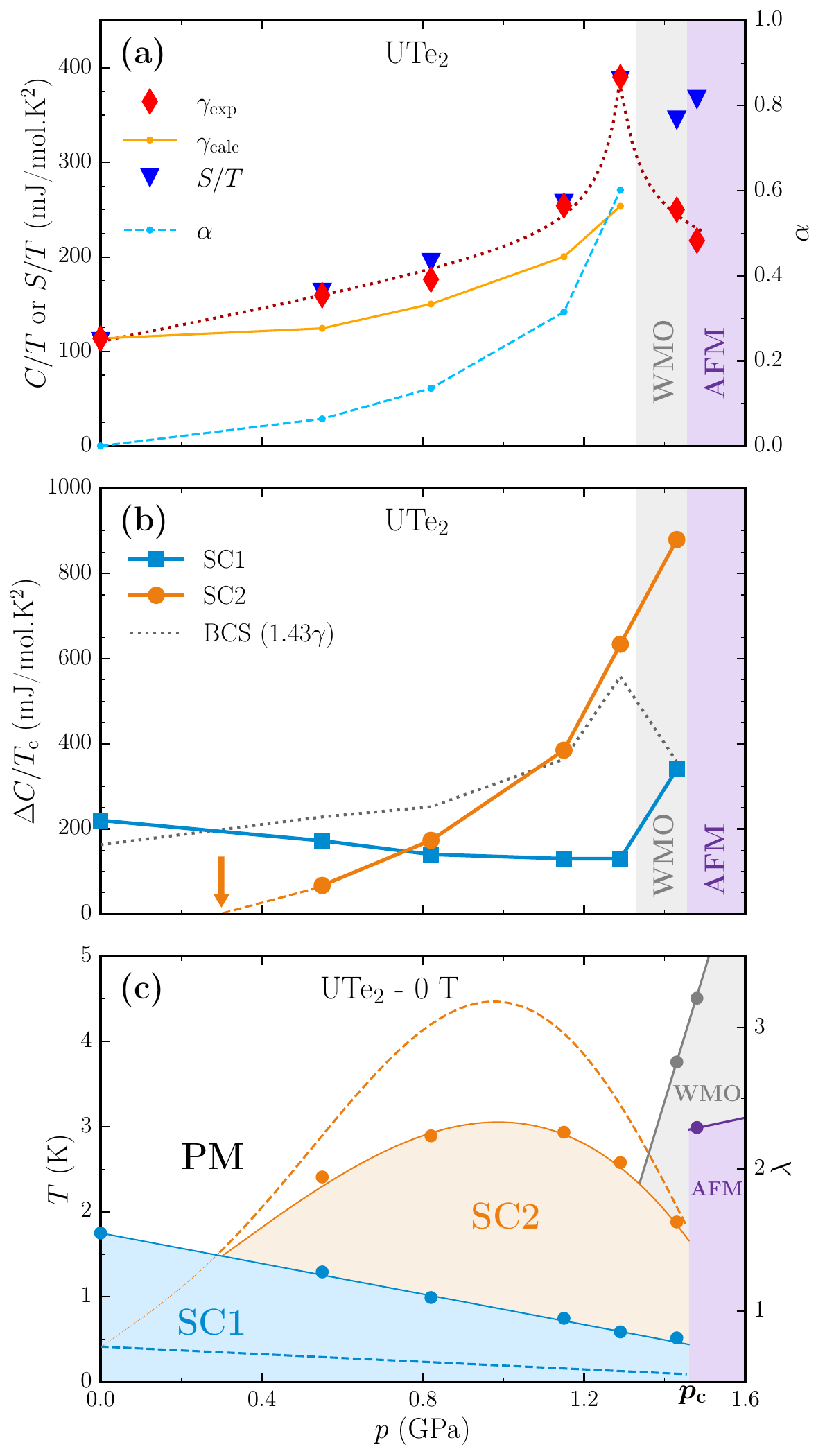}
\caption{\label{fig:compil}(a) Relative evolution of gamma coefficient and entropy as a function of pressure. The red dotted line is a guide for the eye, the orange line a prediction from a strong coupling model with an increase of the fraction ($\alpha$ - blue dashed line on right axis) of the Fermi surface affected by the pairing mechanism of SC2 (see text) (b) Specific heat jumps for both superconducting transitions. The BCS jump expected from gamma is also shown in grey. The orange arrow shows the extrapolated crossing point of SC1 and SC2. (c) Zero-field phase diagram obtained from Fig.~\ref{fig:raw}. Dashed lines: strong coupling constants  (right axis) reproducing the pressure dependence of the critical temperatures of SC1 and SC2.}
\end{figure}

Coming back to the absolute measurements of Fig.~\ref{fig:raw} , we extract the normal-state electronic specific heat and the pressure evolution of the Sommerfeld coefficient $\gamma$. The latter is obtained from the constant value of $C/T$ in the paramagnetic Fermi-liquid regime above the ordered phases, after subtraction of a weak phonon contribution determined at ambient pressure and assumed to be pressure-independent. Further details of the analysis are given in \cite{suppl}). This is shown in Fig.~\ref{fig:compil}(a). The electronic specific heat increases with pressure and reaches a maximum at $\SI{1.29}{\giga\pascal}$, where the Sommerfeld coefficient $\gamma$ is enhanced by a factor of three. This behavior closely parallels the field-induced enhancement of $\gamma$ for $\bm{H}\parallel\bm{b}$, supporting a correlation between the mass renormalization and the reinforcement of pairing in the SC2 phase under both tuning parameters. At the two highest pressures, the determination of $\gamma$ is complicated by the emergence of the WMO and AFM phases. Assuming that the paramagnetic Fermi-liquid regime is recovered at $T=\SI{6}{\kelvin}$, we extract $\gamma$ from the corresponding $C/T$ value and find that it decreases markedly with further increasing pressure.

In Fig.~\ref{fig:compil}(a), we also plot the entropy, calculated by integrating the electronic contribution of $C/T$ (after subtraction of the phonon contribution) up to a temperature in the normal state. This was usually 4K except for the highest two pressures where we used 6K to be above the WMO and AFM transition temperatures. At ambient pressure, the entropy balance at the superconducting transition is well respected, with $S/T = \gamma  \approx \SI{120}{\milli\joule\per(\mole\cdot\kelvin\squared)}$. When increasing the pressure, both $\gamma$ and the entropy rise in a similar fashion, respecting entropy balance even under pressure up to $\SI{1.29}{\giga\pascal}$. For the two points at higher pressure, there is clearly an excess of entropy compared to the value of $\gamma$ determined above $\SI{6}{\kelvin}$, but in the presence of the WMO and AFM phases, there could be additional contributions to entropy.

For both superconducting transitions, the associated jump size in specific heat is plotted in Fig.~\ref{fig:compil}(b) as a function of pressure. We also show the expected BCS value $\Delta C /T_\text{c} (p) = 1.43 \gamma (p)$, taking into account the evolution of $\gamma$ with pressure. At ambient pressure, the specific heat jump size of SC1 is slightly higher than the BCS value. Indeed, ambient pressure measurements on high-quality single crystals have shown that UTe$_2$ is in the strong coupling regime, with $\Delta C/T_\text{c} \approx 2.3 \gamma $ \cite{aoki_molten_2024}. In this study, the jump size appears smaller compared to measurements on Molten Salt Flux (MSF)-grown crystals, due to the lower quality of CVT-grown crystals and the lower $T_\text{c1}$. The jump size and  $T_\text{c1}$ are well correlated \cite{Cairns2020, Rosa2022a}. As pressure increases, the jump associated with SC1 decreases. 
Regarding SC2, an extrapolation to low pressures indicates that the jump size is close to zero at the pressure where the SC2 transition line meets the SC1 boundary (marked with an orange arrow), in agreement with previous studies \cite{Braithwaite2019, vasina_connecting_2025}. 

With increasing pressure, the jump associated with SC2 rapidly grows up to $\SI{1.3}{\giga\pascal}$, like the electronic specific heat coefficient but in contrast to the behavior of $T_\text{c2}$, which has a maximum at around 1 GPa. Finally, at $\SI{1.43}{\giga\pascal}$, the highest pressure where superconductivity is still present in this study, both the SC1 and the SC2 phases show a sudden increase in their jump size even though $\gamma$ decreases. 
At this pressure, where the SC2 phase emerges in the presence of WMO, we also notice a change in the shape of both transitions. 
The SC1 transition gets  extremely sharp with an enhanced jump, while the SC2 anomaly increases too but evolves into a broader, almost symmetrical peak. Large pressure inhomogeneities are ruled out by the sharpness of the SC1 transition, although, as we are close to the pressure where superconductivity is abruptly suppressed, small inhomogeneities may have a strong effect on SC2. A recent ultrasound attenuation study proposed a surprisingly large extension of the superconducting fluctuation regime under pressure near $\SI{0.77}{\giga\pascal}$ for the SC2 transition \cite{kamat2026}. But we do not find no a similar effect in our specific heat data in this pressure range. On the contrary, at such pressures the specific heat anomaly is sharper than at lower pressures, as found already in previous work \cite{vasina_connecting_2025}. A significant broadening and change of behavior of the specific heat anomaly is only observed in a narrow pressure range around $p_\text{c}$, 
and will be further discussed below.
 
 Focusing again on the pressure range below $\SI{1.4}{\giga\pascal}$, where the SC2 phase emerges from a normal state without WMO, the value of $\gamma(p)$, of the specific heat jumps, and the critical temperatures should be related. We can use a simple model where the evolution of $T_\text{c}$ under pressure of SC2 and SC1 is only governed by that of the pairing strength of their respective mechanism (see Supplemental Material \cite{suppl} for more details). 
The large specific heat jump of the SC1 phase at ambient pressure and the strong increase of the Sommerfeld coefficient under pressure point to a strong coupling regime for both superconducting phases. In Fig.~\ref{fig:compil}(c), we have shown a typical pressure dependence of the strong coupling constants $\lambda_{\text{SC2}}$ and $\lambda_{\text{SC1}}$ reproducing $T_\text{c}(p)$  for both phases. 
Strong coupling effects should then lead to a renormalization of the electronic specific heat coefficient as $C/T = \gamma(p) = \gamma_0 [1+  \lambda_{\text{SC1}}(p)][1+ \lambda_{\text{SC2}}(p)]$. This predicts a smooth maximum around $\SI{1}{\giga\pascal}$, far from matching the experimental $\gamma(p)$ reported on Fig.~\ref{fig:compil}(a) \cite{suppl}.
 Moreover, in such a simple strong coupling picture, the specific heat jump of SC2 should always be larger than the BCS jump, whereas 
 the amplitude of $\Delta C/T$ for SC2 seemingly vanishes at low pressure. To account for these discrepancies we suggest a model where the SC2 phase nucleates only on a small fraction of the Fermi surface.
 We can take the data of the absolute specific heat jump as measure of this fraction ($\alpha$) of the Fermi surface \cite{suppl}. 
 Then, renormalization of 
 $\gamma(p)$ by $\lambda_{\text{SC2}}$ should happen only on that fraction of the Fermi surface on which SC2 condenses.
 We report on Fig.~\ref{fig:compil}(a) the resulting behavior calculated for $\alpha(p)$ and $\gamma(p)$ (more details in \cite{suppl}): 
 the prediction for the pressure dependence of $\gamma(p)$, and notably the continuous increase beyond the pressure of the maximum of $T_\text{c}$ for SC2, is much closer to the experimental data. 
 
 Everything happens as if indeed, the SC2 phase first nucleates on a small portion of the Fermi surface, rapidly increasing under pressure. This description can explain how the critical temperature of SC2 decreases at $\SI{1.29}{\giga\pascal}$ whereas $\Delta C/T$ and $\gamma$ increase. It also accounts for the vanishing value of  $\Delta C/T$ for SC2 where the SC1 and SC2 lines meet. This fact is noteworthy, as this condition is required for a thermodynamically consistent phase diagram in which three second-order transition lines meet at a single point \cite{Braithwaite2019, Wu2025PRL}. 
 What this empirical analysis does not solve, is the puzzle of a large strong coupling constant ($\lambda_{\text{SC2}}(p) \geq 1.5$) for the SC2 phase when it emerges near $\SI{0.3}{\giga\pascal}$, with a vanishing weight on the Fermi surface: 
 superconductivity with very small carrier numbers exist in other systems, but it requires very specific, often elusive, pairing mechanisms \cite{CollignonARCMP2019, KimSciAdv2018}.

 Regarding the pressure where SC2 emerges from a WMO phase, the strong increase in the size of the jump in $C/T$ at both the SC2 and SC1 transitions at $\SI{1.43}{\giga\pascal}$ whereas $\gamma$ and $T_\text{c}$ are decreasing, might appear surprising. A possible explanation is suggested by the large entropy measured at 6\,K around the critical pressure: it points to the presence of strong magnetic fluctuations in the WMO phase. Then, the gap opening on the Fermi surface induced by the transition into the superconducting state might also trigger magnetic entropy reduction, hence a larger specific heat jump than predicted even by strong coupling scenarios. 
 If, as our empirical model suggests, SC2 concerns only a portion of the Fermi surface even close to the critical pressure (see Fig.~\ref{fig:compil}(a)),   
further reduction of the magnetic entropy should occur at the SC1 phase transition (taking place on the whole Fermi surface), explaining also its large jump at $\SI{1.43}{\giga\pascal}$ (Fig.~\ref{fig:compil}(b)).

Microscopically, the existence of a maximum under pressure where $\gamma$ is increased by a factor of three gives a direct indication of enhanced electronic correlations, and possible quantum criticality under pressure. Previously, this had been indirectly inferred mainly from transport measurements: the $A$ coefficient of the quadratic Fermi-liquid temperature dependence ($\rho$=$\rho_0+AT^2$) was found to increase by a factor of six \cite{Braithwaite2019} to eight \cite{Valiska2021}. Although the value of the $A$ coefficient depends critically on the current direction \cite{Eo2022, Thebault2022, Knebel2024}, the usual relationship ($m^* \propto A^2$) is compatible with our direct determination of a threefold enhancement of $\gamma$. Another study, analyzing the resistivity with a non-Fermi-liquid law ($\rho = \rho_0 + A_nT^n$), found a minimum of the resistivity exponent $n \approx 1$ under pressure, indicating possible quantum criticality. Microscopic evidence of the increase of AFM spin fluctuations with pressure has been obtained by NMR measurements under pressure \cite{Kinjo2023PRB, Ambika2026}, which suggests the importance of these fluctuations for the development of the SC2 phase. 
A striking aspect of our study is that  $\gamma$, and thus the electronic correlations that are associated with it, reaches its highest value at a pressure significantly lower than the critical pressure $p_\textit{c}$ where long-range AFM order is established. This quite surprising fact was already suggested from the previous transport studies where the maximum of the $A$ coefficient \cite{Braithwaite2019}, and the minimum of the exponent $n$ \cite{Thomas2020}, were also found at pressures below the AFM critical pressure, possibly coinciding with an extrapolated critical pressure for the WMO phase. Indeed, there appears to be a near coincidence of the three phenomena, the maximum of $\gamma$, the maximum in $T_\text{c}$ of SC2, and the extrapolated critical pressure of the WMO phase $p_\text{WMO}$. This could indicate a significant role of the WMO phase in stabilizing SC2 through a quantum critical point. This is not incompatible with a simultaneous competition between the two phases as suggested by our scenario for the enhanced specific heat jumps at $\SI{1.43}{\giga\pascal}$. Indeed superconductivity is known to sometimes suppress the ordered phase whose quantum critical fluctuations are believed to mediate pairing in AFM systems \cite{Knebel2011} and charge density wave systems \cite{Fradkin2015, Chen_2016}.

While this, and several other points, remain to be confirmed or clarified, our quantitative calorimetry study of UTe$_2$ under pressure brings solid new elements to the previous qualitative studies, determining precisely the threefold enhancement of the electronic effective mass with pressure, and quantitative analysis of the pressure effects on the height of the $C/T$ jumps at the superconducting transitions. The direct measurement of a maximum of $\gamma$ under pressure, seemingly linked to 
the emergence of the WMO phase above SC2, and the maximum of the SC2 critical temperature, possibly linked to the WMO critical pressure, 
provide evidence of a strong interplay between the WMO and the high-pressure superconducting phase.

\begin{acknowledgements}
    We thank D. Aoki, M. Houzet and M. Zhitomirsky for fruitful discussions. This work was supported by the French National Agency for Research (ANR) within the projects  FRESCO No. ANR-20-CE30-0020 and SCATE No. ANR-22-CE30-0040. 
    The quantitative heat capacity experiments under hydrostatic pressure were performed at the Max Planck Institute for Chemical Physics of Solids (MPI CPfS) in Dresden, Germany. T.V. thanks MPI CPfS for their hospitality. E. H. acknowledges funding by the DFG through CRC1143 (Project No. 247310070) and the W\"{u}rzburg- Dresden Cluster of Excellence on Complexity and Topology in Quantum Matter—ct.qmat (EXC 2147, Project 6 ID 390858490) and by the ERC Consolidator Grant Ixtreme (GA 101125759). 
 \end{acknowledgements}

\bibliographystyle{apsrev4-2}
\bibliography{ute2}

@Article{Khim2021,
  author        = {Khim, S. and Landaeta, J. F. and Banda, J. and Bannor, N. and Brando, M. and Brydon, P. M. R. and Hafner, D. and Kuechler, R. and Cardoso-Gil, R. and Stockert, U. and Mackenzie, A. P. and Agterberg, D. F. and Geibel, C. and Hassinger, E.},
  journal       = {Science},
  title         = {{Field-induced transition within the superconducting state of CeRh2As2}},
  year          = {{2021}},
  issn          = {{0036-8075}},
  month         = {{AUG 27}},
  number        = {{6558}},
  pages         = {{1012+}},
  volume        = {{373}},
  doi           = {{10.1126/science.abe7518}},
  eissn         = {{1095-9203}},
  orcid-numbers = {{Brydon, Philip/0000-0001-8854-0623 Landaeta, Javier/0000-0002-4229-2115}},
  unique-id     = {{WOS:000690202600034}},
}

@Article{Aoki2019,
  author  = {Aoki ,Dai and Nakamura ,Ai and Honda ,Fuminori and Li ,DeXin and Homma ,Yoshiya and Shimizu ,Yusei and Sato ,Yoshiki J. and Knebel ,Georg and Brison ,Jean-Pascal and Pourret ,Alexandre and Braithwaite ,Daniel and Lapertot ,Gerard and Niu ,Qun and Vali\v{s}ka ,Michal and Harima ,Hisatomo and Flouquet ,Jacques},
  journal = {J. Phys. Soc. Jpn.},
  title   = {{Unconventional Superconductivity in Heavy Fermion UTe$_2$}},
  year    = {2019},
  number  = {4},
  pages   = {043702},
  volume  = {88},
  doi     = {10.7566/JPSJ.88.043702},
  url     = {https://doi.org/10.7566/JPSJ.88.043702},
}

@Article{Ran2019,
  author    = {Ran, Sheng and Eckberg, Chris and Ding, Qing-Ping and Furukawa, Yuji and Metz, Tristin and Saha, Shanta R. and Liu, I-Lin and Zic, Mark and Kim, Hyunsoo and Paglione, Johnpierre and Butch, Nicholas P.},
  journal   = {Science},
  title     = {Nearly ferromagnetic spin-triplet superconductivity},
  year      = {2019},
  issn      = {0036-8075},
  number    = {6454},
  pages     = {684--687},
  volume    = {365},
  doi       = {10.1126/science.aav8645},
  publisher = {American Association for the Advancement of Science},
  url       = {https://science.sciencemag.org/content/365/6454/684},
}

@Article{Ran2019a,
  author   = {Ran, Sheng and Liu, I.-Lin and Eo, Yun Suk and Campbell, Daniel J. and Neves, Paul M. and Fuhrman, Wesley T. and Saha, Shanta R. and Eckberg, Christopher and Kim, Hyunsoo and Graf, David and Balakirev, Fedor and Singleton, John and Paglione, Johnpierre and Butch, Nicholas P.},
  title    = {Extreme magnetic field-boosted superconductivity},
  journal  = {Nat. Phys.},
  year     = {2019},
  volume   = {15},
  pages    = {1250-1254},
  month    = oct,
  issn     = {1745-2481},
  refid    = {Ran2019},
  url      = {https://doi.org/10.1038/s41567-019-0670-x},
}

@Article{Braithwaite2019,
  author   = {Braithwaite, D. and Vali{\v{s}}ka, M. and Knebel, G. and Lapertot, G. and Brison, J. P. and Pourret, A. and Zhitomirsky, M. E. and Flouquet, J. and Honda, F. and Aoki, D.},
  journal  = {Commun Phys},
  title    = {{Multiple superconducting phases in a nearly ferromagnetic system}},
  year     = {2019},
  issn     = {23993650},
  number   = {1},
  volume   = {2},
  doi      = {10.1038/s42005-019-0248-z},
  isbn     = {4200501902},
}

@Article{Ran2020,
  author    = {Ran, Sheng and Kim, Hyunsoo and Liu, I-Lin and Saha, Shanta R and Hayes, Ian and Metz, Tristin and Eo, Yun Suk and Paglione, Johnpierre and Butch, Nicholas P},
  journal   = {Phys. Rev. B},
  title     = {{Enhancement and reentrance of spin triplet superconductivity in UTe$_2$ under pressure}},
  year      = {2020},
  month     = {apr},
  number    = {14},
  pages     = {140503},
  volume    = {101},
  doi       = {10.1103/PhysRevB.101.140503},
  publisher = {American Physical Society},
  url       = {https://link.aps.org/doi/10.1103/PhysRevB.101.140503},
}

@Article{Aoki2020,
  author        = {Aoki, Dai and Honda, Fuminori and Knebel, Georg and Braithwaite, Daniel and Nakamura, Ai and Li, De Xin and Homma, Yoshiya and Shimizu, Yusei and Sato, Yoshiki J. and Brison, Jean Pascal and Flouquet, Jacques},
  journal       = {J. Phys. Soc. Jpn.},
  title         = {{Multiple superconducting phases and unusual enhancement of the upper critical field in UTe$_2$}},
  year          = {2020},
  issn          = {13474073},
  number        = {5},
  pages         = {1--5},
  volume        = {89},
  archiveprefix = {arXiv},
  arxivid       = {2003.09782},
  doi           = {10.7566/JPSJ.89.053705},
}

@Article{Knebel2020,
  author        = {Knebel, Georg and Kimata, Motoi and Vali{\v{s}}ka, Michal and Honda, Fuminori and Li, De Xin and Braithwaite, Daniel and Lapertot, G{\'{e}}rard and Knafo, William and Pourret, Alexandre and Sato, Yoshiki J. and Shimizu, Yusei and Kihara, Takumi and Brison, Jean Pascal and Flouquet, Jacques and Aoki, Dai},
  journal       = {J. Phys. Soc. Jpn.},
  title         = {{Anisotropy of the upper critical field in the heavy-fermion superconductor UTe$_2$ under pressure}},
  year          = {2020},
  issn          = {13474073},
  number        = {5},
  pages         = {1--5},
  volume        = {89},
  archiveprefix = {arXiv},
  arxivid       = {2003.08728},
  doi           = {10.7566/JPSJ.89.053707},
  url           = {https://doi.org/10.7566/JPSJ.89.053707},
}

@Article{Aoki2021,
  author  = {Aoki ,Dai and Kimata ,Motoi and Sato ,Yoshiki J. and Knebel ,Georg and Honda ,Fuminori and Nakamura ,Ai and Li ,Dexin and Homma ,Yoshiya and Shimizu ,Yusei and Knafo ,William and Braithwaite ,Daniel and Vališka ,Michal and Pourret ,Alexandre and Brison ,Jean-Pascal and Flouquet ,Jacques},
  journal = {J. Phys. Soc. Jpn.},
  title   = {{Field-Induced Superconductivity near the Superconducting Critical Pressure in UTe$_2$}},
  year    = {2021},
  number  = {7},
  pages   = {074705},
  volume  = {90},
  doi     = {10.7566/JPSJ.90.074705},
  url     = {https://doi.org/10.7566/JPSJ.90.074705},
}

@Article{Cairns2020,
  author   = {Cairns, Luke Pritchard and Stevens, Callum R. and O'Neill, Christopher D and Huxley, Andrew},
  journal  = {J. Phys.: Condens. Matter},
  title    = {{Composition dependence of the superconducting properties of UTe$_2$}},
  year     = {2020},
  issn     = {0953-8984},
  month    = {sep},
  number   = {41},
  pages    = {415602},
  volume   = {32},
  doi      = {10.1088/1361-648X/ab9c5d},
  url      = {https://iopscience.iop.org/article/10.1088/1361-648X/ab9c5d},
}

@Article{Li2021,
  author  = {Li ,Dexin and Nakamura ,Ai and Honda ,Fuminori and Sato ,Yoshiki J. and Homma ,Yoshiya and Shimizu ,Yusei and Ishizuka ,Jun and Yanase ,Youichi and Knebel ,Georg and Flouquet ,Jacques and Aoki ,Dai},
  journal = {J. Phys. Soc. Jpn.},
  title   = {{Magnetic Properties under Pressure in Novel Spin-Triplet Superconductor UTe$_2$}},
  year    = {2021},
  number  = {7},
  pages   = {073703},
  volume  = {90},
  doi     = {10.7566/JPSJ.90.073703},
  url     = {https://doi.org/10.7566/JPSJ.90.073703},
}

@Misc{suppl,
  note = {In the Supplemental Material we (1) report experimental details on the quantiative calorimetry, (2) indicate the determination of $\gamma,S, T_\text{c}$, and $\Delta C$ as a function of pressure, (3) show additional data obtained on a MSF sample in a diamond anvil cell. [URL]},
}

@article{ajeesh2024,
author = {Ajeesh ,M. O. and Thompson ,Joe D. and Bauer ,Eric D. and Ronning ,Filip and Thomas ,Sean M. and Rosa ,Priscila F. S.},
title = {Hydrostatic Pressure Studies on Non-Superconducting UTe2},
journal = {J. Phys. Soc. Jpn.},
volume = {93},
number = {5},
pages = {055001},
year = {2024},
doi = {10.7566/JPSJ.93.055001},
URL = {https://doi.org/10.7566/JPSJ.93.055001}
}

@article{honda_pressure-induced_2023,
	title = {Pressure-induced Structural Phase Transition and New Superconducting Phase in {UTe}$_{2}$},
	volume = {92},
	issn = {0031-9015, 1347-4073},
	url = {https://journals.jps.jp/doi/10.7566/JPSJ.92.044702},
	doi = {10.7566/JPSJ.92.044702},
	pages = {044702},
	number = {4},
	journal = {J. Phys. Soc. Jpn.},
	author = {Honda, Fuminori and Kobayashi, Shintaro and Kawamura, Naomi and Kawaguchi, Saori I. and Koizumi, Takatsugu and Sato, Yoshiki J. and Homma, Yoshiya and Ishimatsu, Naoki and Gouchi, Jun and Uwatoko, Yoshiya and Harima, Hisatomo and Flouquet, Jacques and Aoki, Dai},
	urldate = {2024-06-25},
	date = {2023-04-15},
    year={2024},
	langid = {english},
}

@Article{Sauls1994,
  author   = {J.A. Sauls},
  journal  = {Adv. Phys.},
  title    = {The order parameter for the superconducting phases of UPt3},
  year     = {1994},
  number   = {1},
  pages    = {113-141},
  volume   = {43},
  doi      = {10.1080/00018739400101475},
  eprint   = {http://dx.doi.org/10.1080/00018739400101475},
}

@Article{Ott1985,
  author    = {Ott, H. R. and Rudigier, H. and Fisk, Z. and Smith, J. L.},
  title     = {Phase transition in the superconducting state of ${\mathrm{U}}_{1\mathrm{\ensuremath{-}}\mathrm{x}}$${\mathrm{Th}}_{\mathrm{x}}$${\mathrm{Be}}_{13}$ (x=0--0.06)},
  journal   = {Phys. Rev. B},
  year      = {1985},
  volume    = {31},
  pages     = {1651--1653},
  month     = {Feb},
  doi       = {10.1103/PhysRevB.31.1651},
  issue     = {3},
  numpages  = {0},
  publisher = {American Physical Society},
  url       = {https://link.aps.org/doi/10.1103/PhysRevB.31.1651},
}

@Article{Valiska2021,
  author    = {Vali\ifmmode \check{s}\else \v{s}\fi{}ka, M. and Knafo, W. and Knebel, G. and Lapertot, G. and Aoki, D. and Braithwaite, D.},
  journal   = {Phys. Rev. B},
  title     = {{Magnetic reshuffling and feedback on superconductivity in ${\mathrm{UTe}}_{2}$ under pressure}},
  year      = {2021},
  month     = {Dec},
  pages     = {214507},
  volume    = {104},
  doi       = {10.1103/PhysRevB.104.214507},
  issue     = {21},
  numpages  = {11},
  publisher = {American Physical Society},
  url       = {https://link.aps.org/doi/10.1103/PhysRevB.104.214507},
}

@Article{Rosuel2023,
  author        = {Rosuel, Adrien and Marcenat, Christophe and Knebel, Georg and Klein, Thierry and Pourret, Alexandre and Marquardt, Nils and Niu, Qun and Rousseau, Simon and Demuer, Albin and Seyfarth, Gabriel and Lapertot, G{\'{e}}rard and Aoki, Dai and Braithwaite, Daniel and Flouquet, Jacques and Brison, J. P.},
  journal       = {Phys. Rev. X},
  title         = {{Field-Induced Tuning of the Pairing State in a Superconductor}},
  year          = {2023},
  issn          = {2160-3308},
  month         = {feb},
  number        = {1},
  pages         = {011022},
  volume        = {13},
  archiveprefix = {arXiv},
  arxivid       = {2205.04524},
  doi           = {10.1103/PhysRevX.13.011022},
  url           = {http://arxiv.org/abs/2205.04524 http://dx.doi.org/10.1103/PhysRevX.13.011022 https://link.aps.org/doi/10.1103/PhysRevX.13.011022},
}

@Article{Thebault2022,
  author    = {Thebault, T. and Vali\ifmmode \check{s}\else \v{s}\fi{}ka, M. and Lapertot, G. and Pourret, A. and Aoki, D. and Knebel, G. and Braithwaite, D. and Knafo, W.},
  journal   = {Phys. Rev. B},
  title     = {{Anisotropic signatures of electronic correlations in the electrical resistivity of ${\mathrm{UTe}}_{2}$}},
  year      = {2022},
  month     = {Oct},
  pages     = {144406},
  volume    = {106},
  doi       = {10.1103/PhysRevB.106.144406},
  issue     = {14},
  numpages  = {14},
  publisher = {American Physical Society},
  url       = {https://link.aps.org/doi/10.1103/PhysRevB.106.144406},
}

@Article{Eo2022,
  author    = {Eo, Yun Suk and Liu, Shouzheng and Saha, Shanta R. and Kim, Hyunsoo and Ran, Sheng and Horn, Jarryd A. and Hodovanets, Halyna and Collini, John and Metz, Tristin and Fuhrman, Wesley T. and Nevidomskyy, Andriy H. and Denlinger, Jonathan D. and Butch, Nicholas P. and Fuhrer, Michael S. and Wray, L. Andrew and Paglione, Johnpierre},
  journal   = {Phys. Rev. B},
  title     = {$c$-axis transport in {${\mathrm{UTe}}_{2}$}: Evidence of three-dimensional conductivity component},
  year      = {2022},
  month     = {Aug},
  pages     = {L060505},
  volume    = {106},
  doi       = {10.1103/PhysRevB.106.L060505},
  issue     = {6},
  numpages  = {7},
  publisher = {American Physical Society},
  url       = {https://link.aps.org/doi/10.1103/PhysRevB.106.L060505},
}

@Article{Tokunaga2023,
  author    = {Tokunaga, Y. and Sakai, H. and Kambe, S. and Opletal, P. and Tokiwa, Y. and Haga, Y. and Kitagawa, S. and Ishida, K. and Aoki, D. and Knebel, G. and Lapertot, G. and Kr\"amer, S. and Horvati\ifmmode \acute{c}\else \'{c}\fi{}, M.},
  journal   = {Phys. Rev. Lett.},
  title     = {{Longitudinal Spin Fluctuations Driving Field-Reinforced Superconductivity in ${\mathrm{UTe}}_{2}$}},
  year      = {2023},
  month     = {Nov},
  pages     = {226503},
  volume    = {131},
  doi       = {10.1103/PhysRevLett.131.226503},
  issue     = {22},
  numpages  = {7},
  publisher = {American Physical Society},
  url       = {https://link.aps.org/doi/10.1103/PhysRevLett.131.226503},
}

@article{knafo_incommensurate_2025,
	title = {Incommensurate Antiferromagnetism in {UTe}$_{2}$ under Pressure},
	volume = {15},
	issn = {2160-3308},
	url = {https://link.aps.org/doi/10.1103/PhysRevX.15.021075},
	doi = {10.1103/PhysRevX.15.021075},
	pages = {021075},
	number = {2},
	journal = {Phys. Rev. X},
	author = {Knafo, W. and Thebault, T. and Raymond, S. and Manuel, P. and Khalyavin, D. D. and Orlandi, F. and Ressouche, E. and Beauvois, K. and Lapertot, G. and Kaneko, K. and Aoki, D. and Braithwaite, D. and Knebel, G.},
	urldate = {2025-08-05},
    year = {2025},
	date = {2025-05-30},
	langid = {english},
}

@article{aoki_molten_2024,
	title = {Molten Salt Flux Liquid Transport Method for Ultra Clean Single Crystals {UTe}$_{\textrm{2}}$},
	volume = {93},
	issn = {0031-9015, 1347-4073},
	url = {https://journals.jps.jp/doi/10.7566/JPSJ.93.043703},
	doi = {10.7566/JPSJ.93.043703},
	pages = {043703},
	number = {4},
	journal = {J. Phys. Soc. Jpn.},
	author = {Aoki, Dai},
	urldate = {2025-08-25},
	date = {2024-04-15},
    year = {2024},
}

@Article{Kinjo2023PRB,
  author    = {Kinjo, K. and Fujibayashi, H. and Kitagawa, S. and Ishida, K. and Tokunaga, Y. and Sakai, H. and Kambe, S. and Nakamura, A. and Shimizu, Y. and Homma, Y. and Li, D. X. and Honda, F. and Aoki, D. and Hiraki, K. and Kimata, M. and Sasaki, T.},
  journal   = {Phys. Rev. B},
  title     = {{Change of superconducting character in ${\mathrm{UTe}}_{2}$ induced by magnetic field}},
  year      = {2023},
  month     = {Feb},
  pages     = {L060502},
  volume    = {107},
  doi       = {10.1103/PhysRevB.107.L060502},
  issue     = {6},
  numpages  = {5},
  publisher = {American Physical Society},
  url       = {https://link.aps.org/doi/10.1103/PhysRevB.107.L060502},
}

@Article{Thomas2020,
  author    = {Thomas, S. M. and Santos, F. B. and Christensen, M. H. and Asaba, T. and Ronning, F. and Thompson, J. D. and Bauer, E. D. and Fernandes, R. M. and Fabbris, G. and Rosa, P. F. S.},
  journal   = {Sci. Adv.},
  title     = {{Evidence for a pressure-induced antiferromagnetic quantum critical point in intermediate-valence UTe$_2$}},
  year      = {2024},
  month     = jan,
  number    = {42},
  pages     = {eabc8709},
  volume    = {6},
  comment   = {doi: 10.1126/sciadv.abc8709},
  doi       = {10.1126/sciadv.abc8709},
  publisher = {American Association for the Advancement of Science},
  url       = {https://doi.org/10.1126/sciadv.abc8709},
}

@Article{Knebel2024,
  author    = {Knebel, G. and Pourret, A. and Rousseau, S. and Marquardt, N. and Braithwaite, D. and Honda, F. and Aoki, D. and Lapertot, G. and Knafo, W. and Seyfarth, G. and Brison, J-P. and Flouquet, J.},
  journal   = {Phys. Rev. B},
  title     = {$c$-axis electrical transport at the metamagnetic transition in the heavy-fermion superconductor {UTe$_2$} under pressure},
  year      = {2024},
  issn      = {2469-9969},
  month     = apr,
  number    = {15},
  pages     = {155103},
  volume    = {109},
  doi       = {10.1103/physrevb.109.155103},
  publisher = {American Physical Society (APS)},
}

@article{vasina_connecting_2025,
	title = {Connecting High-Field and High-Pressure Superconductivity in {UTe}$_{2}$},
	volume = {134},
	rights = {All rights reserved},
	issn = {0031-9007, 1079-7114},
	url = {https://link.aps.org/doi/10.1103/PhysRevLett.134.096501},
	doi = {10.1103/PhysRevLett.134.096501},
	pages = {096501},
	number = {9},
    year = {2025},
    journal = {Phys. Rev. Lett.},
	author = {Vasina, T. and Aoki, D. and Miyake, A. and Seyfarth, G. and Pourret, A. and Marcenat, C. and Amano Patino, M. and Lapertot, G. and Flouquet, J. and Brison, J.-P. and Braithwaite, D. and Knebel, G.},
	urldate = {2025-03-24},
	date = {2025-03-04},
	langid = {english},
}

@misc{kamat2026,
      title={Vanishing Phase Stiffness and Fluctuation-Dominated Superconductivity: Evidence for Inter-Band Pairing in UTe$_2$}, 
      author={Sahas Kamat and Jared Dans and Shanta Saha and Daniel F. Agterberg and Johnpierre Paglione and B. J. Ramshaw},
      year={2026},
      eprint={2601.09138},
      archivePrefix={arXiv},
      primaryClass={cond-mat.supr-con},
      url={https://arxiv.org/abs/2601.09138}, 
}

@article{Wu2025,
author = {Wu, Z. and Weinberger, T. I. and Hickey, A. J. and Chichinadze, D. V. and Shaffer, D. and Cabala, A. and Chen, H. and Long, M. and Brumm, T. J. and Xie, W. and Ling, Y. and Zhu, Z. and Skourski, Y. and Graf, D. E. and Sechovsk{\'{y}}, V. and Vali{\v{s}}ka, M. and Lonzarich, G. G. and Grosche, F. M. and Eaton, A. G.},
doi = {10.1103/PhysRevX.15.021019},
issn = {21603308},
journal = {Physical Review X},
number = {2},
pages = {1--20},
title = {{A Quantum Critical Line Bounds the High Field Metamagnetic Transition Surface in UTe2}},
volume = {15},
year = {2025}
}

@Article{Wu2025PRL,
  author    = {Wu, Zheyu and Chen, Jiasheng and Weinberger, Theodore I. and Cabala, Andrej and Sechovský, Vladimír and Vališka, Michal and Alireza, Patricia L. and Eaton, Alexander G. and Grosche, F. Malte},
  journal   = {Physical Review Letters},
  title     = {Magnetic Signatures of Pressure-Induced Multicomponent Superconductivity in <mml:math xmlns:mml="http://www.w3.org/1998/Math/MathML" display="inline"><mml:mrow><mml:msub><mml:mrow><mml:mi>UTe</mml:mi></mml:mrow><mml:mrow><mml:mn>2</mml:mn></mml:mrow></mml:msub></mml:mrow></mml:math>},
  year      = {2025},
  issn      = {1079-7114},
  month     = jun,
  number    = {23},
  pages     = {236501},
  volume    = {134},
  date      = {2025-06-10},
  day       = {10},
  doi       = {10.1103/physrevlett.134.236501},
  publisher = {American Physical Society (APS)},
}

@article{Boyer1983,
title = {The low-temperature specific heat of PTFE (Teflon) at pressures to 5.2 kbar},
journal = {Journal of Non-Crystalline Solids},
volume = {55},
number = {3},
pages = {413-432},
year = {1983},
issn = {0022-3093},
doi = {https://doi.org/10.1016/0022-3093(83)90046-7},
author = {James D. Boyer and J.C. Lasjaunias and R.A. Fisher and Norman E. Phillips}
}

@unpublished{Nicklas_unpublished,
	author = {Michael Nicklas},
	note = {Unpublished data},
	title = {Low-temperature pressure dependence of the specific heat of Flourinert FC-72 up to 1.5 GPa }}

@article{Eiling1981,
	author = {A Eiling and J S Schilling},
	doi = {10.1088/0305-4608/11/3/010},
	journal = {Journal of Physics F: Metal Physics},
	month = {Mar},
	number = {3},
	pages = {623--639},
	publisher = {{IOP} Publishing},
	title = {Pressure and temperature dependence of electrical resistivity of {Pb} and {Sn} from 1-300 {K} and 0-10 {GPa} - use as continuous resistive pressure monitor accurate over wide temperature range superconductivity under pressure in {Pb}, {Sn} and {In}},
	url = {https://doi.org/10.1088/0305-4608/11/3/010},
	volume = {11},
	year = {1981}}

@article{Ambika2026,
   title={Enhancement of antiferromagnetic spin fluctuations in {UTe$_2$} under pressure revealed by
                    {$^{125}$Te}
                    NMR},
   volume={113},
   ISSN={2469-9969},
   url={http://dx.doi.org/10.1103/4gcm-mbl7},
   DOI={10.1103/4gcm-mbl7},
   number={1},
   journal={Physical Review B},
   publisher={American Physical Society (APS)},
   author={Vijayan Ambika, Devi and Ding, Qing-Ping and Frank, Corey E. and Ran, Sheng and Butch, Nicholas P. and Furukawa, Yuji},
   year={2026},
   month=jan }

@Article{Rosa2022a,
  author    = {Rosa, Priscila F. S. and Weiland, Ashley and Fender, Shannon S. and Scott, Brian L. and Ronning, Filip and Thompson, Joe D. and Bauer, Eric D. and Thomas, Sean M.},
  journal   = {Communications Materials},
  title     = {Single thermodynamic transition at 2 K in superconducting UTe2 single crystals},
  year      = {2022},
  issn      = {2662-4443},
  month     = may,
  number    = {1},
  volume    = {3},
  doi       = {10.1038/s43246-022-00254-2},
  publisher = {Springer Science and Business Media LLC},
}

@inBook{Nicklas2015,
author="Nicklas, Michael",
editor="Avella, Adolfo
and Mancini, Ferdinando",
title="Pressure Probes",
bookTitle="Strongly Correlated Systems: Experimental Techniques",
year="2015",
publisher="Springer Berlin Heidelberg",
address="Berlin, Heidelberg",
pages="173--204",
isbn="978-3-662-44133-6",
doi="10.1007/978-3-662-44133-6_6",
url="https://doi.org/10.1007/978-3-662-44133-6_6"
}

@article{Wilhelm2004,
    author = {Wilhelm, H. and Lühmann, T. and Rus, T. and Steglich, F.},
    title = {A compensated heat-pulse calorimeter for low temperatures},
    journal = {Review of Scientific Instruments},
    volume = {75},
    number = {8},
    pages = {2700-2705},
    year = {2004},
    month = {08},
    issn = {0034-6748},
    doi = {10.1063/1.1771486},
    url = {https://doi.org/10.1063/1.1771486},
}

@Article{CollignonARCMP2019,
  author    = {Collignon, Clément and Lin, Xiao and Rischau, Carl Willem and Fauqué, Benoît and Behnia, Kamran},
  journal   = {Annual Review of Condensed Matter Physics},
  title     = {Metallicity and Superconductivity in Doped Strontium Titanate},
  year      = {2019},
  issn      = {1947-5462},
  number    = {Volume 10, 2019},
  pages     = {25-44},
  volume    = {10},
  doi       = {https://doi.org/10.1146/annurev-conmatphys-031218-013144},
  publisher = {Annual Reviews},
  type      = {Journal Article},
  url       = {https://www.annualreviews.org/content/journals/10.1146/annurev-conmatphys-031218-013144},
}

@Article{KimSciAdv2018,
  author   = {Hyunsoo Kim and Kefeng Wang and Yasuyuki Nakajima and Rongwei Hu and Steven Ziemak and Paul Syers and Limin Wang and Halyna Hodovanets and Jonathan D. Denlinger and Philip M. R. Brydon and Daniel F. Agterberg and Makariy A. Tanatar and Ruslan Prozorov and Johnpierre Paglione },
  journal  = {Science Advances},
  title    = {Beyond triplet: Unconventional superconductivity in a spin-3/2 topological semimetal},
  year     = {2018},
  number   = {4},
  pages    = {eaao4513},
  volume   = {4},
  doi      = {10.1126/sciadv.aao4513},
  eprint   = {https://www.science.org/doi/pdf/10.1126/sciadv.aao4513},
  url      = {https://www.science.org/doi/abs/10.1126/sciadv.aao4513},
}

@Article{MarsiglioPRB1986,
  author    = {Marsiglio, F. and Carbotte, J. P.},
  journal   = {Phys. Rev. B},
  title     = {Strong-coupling corrections to Bardeen-Cooper-Schrieffer ratios},
  year      = {1986},
  month     = {May},
  pages     = {6141--6146},
  volume    = {33},
  doi       = {10.1103/PhysRevB.33.6141},
  issue     = {9},
  numpages  = {0},
  publisher = {American Physical Society},
  url       = {https://link.aps.org/doi/10.1103/PhysRevB.33.6141},
}

@Article{TakadaJPSJ1980,
  author   = {Takada ,Yasutami},
  journal  = {Journal of the Physical Society of Japan},
  title    = {Theory of Superconductivity in Polar Semiconductors and Its Application to N-Type Semiconducting SrTiO3},
  year     = {1980},
  number   = {4},
  pages    = {1267-1275},
  volume   = {49},
  doi      = {10.1143/JPSJ.49.1267},
  eprint   = {https://doi.org/10.1143/JPSJ.49.1267},
  url      = {https://doi.org/10.1143/JPSJ.49.1267},
}

@article{Knebel2011,
     author = {Georg Knebel and Dai Aoki and Jacques Flouquet},
     title = {Antiferromagnetism and superconductivity in cerium based heavy-fermion compounds},
     journal = {Comptes Rendus. Physique},
     pages = {542--566},
     year = {2011},
     publisher = {Elsevier},
     volume = {12},
     number = {5-6},
     doi = {10.1016/j.crhy.2011.05.002},
     language = {en},
}

@article{Fradkin2015,
  title = {Colloquium: Theory of intertwined orders in high temperature superconductors},
  author = {Fradkin, Eduardo and Kivelson, Steven A. and Tranquada, John M.},
  journal = {Rev. Mod. Phys.},
  volume = {87},
  issue = {2},
  pages = {457--482},
  numpages = {26},
  year = {2015},
  month = {May},
  publisher = {American Physical Society},
  doi = {10.1103/RevModPhys.87.457},
  url = {https://link.aps.org/doi/10.1103/RevModPhys.87.457}
}

@article{Chen_2016,
doi = {10.1088/0034-4885/79/8/084505},
url = {https://doi.org/10.1088/0034-4885/79/8/084505},
year = {2016},
month = {jul},
publisher = {IOP Publishing},
volume = {79},
number = {8},
pages = {084505},
author = {Chen, Chih-Wei and Choe, Jesse and Morosan, E},
title = {Charge density waves in strongly correlated electron systems},
journal = {Reports on Progress in Physics}
}

\end{document}


\newcommand{\warn}[1]{\textcolor{red}{[#1]}}

\renewcommand{\thefigure}{S\arabic{figure}}

\newcommand{\pheliqs}{Univ. Grenoble Alpes, CEA, Grenoble INP, IRIG, PHELIQS, F-38000 Grenoble, France}
\newcommand{\neel}{Univ. Grenoble Alpes, CNRS, Institut Néel, F-38000 Grenoble, France}
\newcommand{\lncmiG}{Univ. Grenoble Alpes, INSA Toulouse, Univ. Toulouse Paul Sabatier, EMFL, CNRS, LNCMI, F-38000 Grenoble, France}
\newcommand{\imr}{Institute for Materials Research, Tohoku University, Oarai, Ibaraki, 311-1313, Japan}
\newcommand{\TU}{Institute for Solid State and Materials Physics, TU Dresden University of Technology, 01062 Dresden, Germany}
\newcommand{\MPG}{Max Planck Institute for Chemical Physics of Solids, 01187 Dresden, Germany}
\newcommand{\KIT}{Institute for Quantum Materials and Technologies, Karlsruhe Institute of Technology, Kaiserstra\ss e 12, 76131 Karlsruhe, Germany }

\title{Supplemental materials to the manuscript \\ "Quantitative thermodynamic study of superconducting and normal states  in \texorpdfstring{UTe$_2$}{UTe2} under pressure"}

\author{T.~Vasina}
\thanks{These authors contributed equally to this work.}
\affiliation{\pheliqs}
\author{M.~Pfeiffer}
\thanks{These authors contributed equally to this work.}
\affiliation{\KIT}
\affiliation{\MPG}
\affiliation{\TU}
\author{R.~Borth}
\affiliation{\MPG}
\author{M.~Nicklas}
\affiliation{\MPG}
\author{M.~Amano Patino}
\affiliation{\pheliqs}
\affiliation{\neel}
\author{G.~Lapertot}
\affiliation{\pheliqs}
\author{J.-P.~Brison}
\affiliation{\pheliqs}
\author{E.~Hassinger}
\affiliation{\KIT}
\affiliation{\MPG}
\author{G.~Knebel}
\email[E-mail me at: ]{georg.knebel@cea.fr}
\affiliation{\pheliqs}
\author{D.~Braithwaite}
\email[E-mail me at: ]{daniel.braithwaite@cea.fr}
\affiliation{\pheliqs}

\maketitle

\section{Experimental details}
\subsection{CVT sample}
\begin{figure}[ht]
    \centering
    \raisebox{0.35\height}{
    \includegraphics[width=0.38\linewidth]{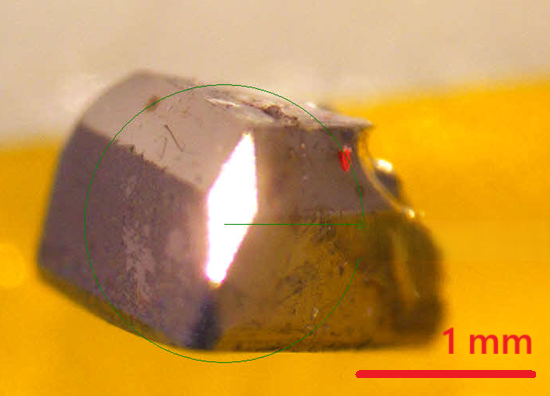}}\quad
    \includegraphics[width=0.55\linewidth]{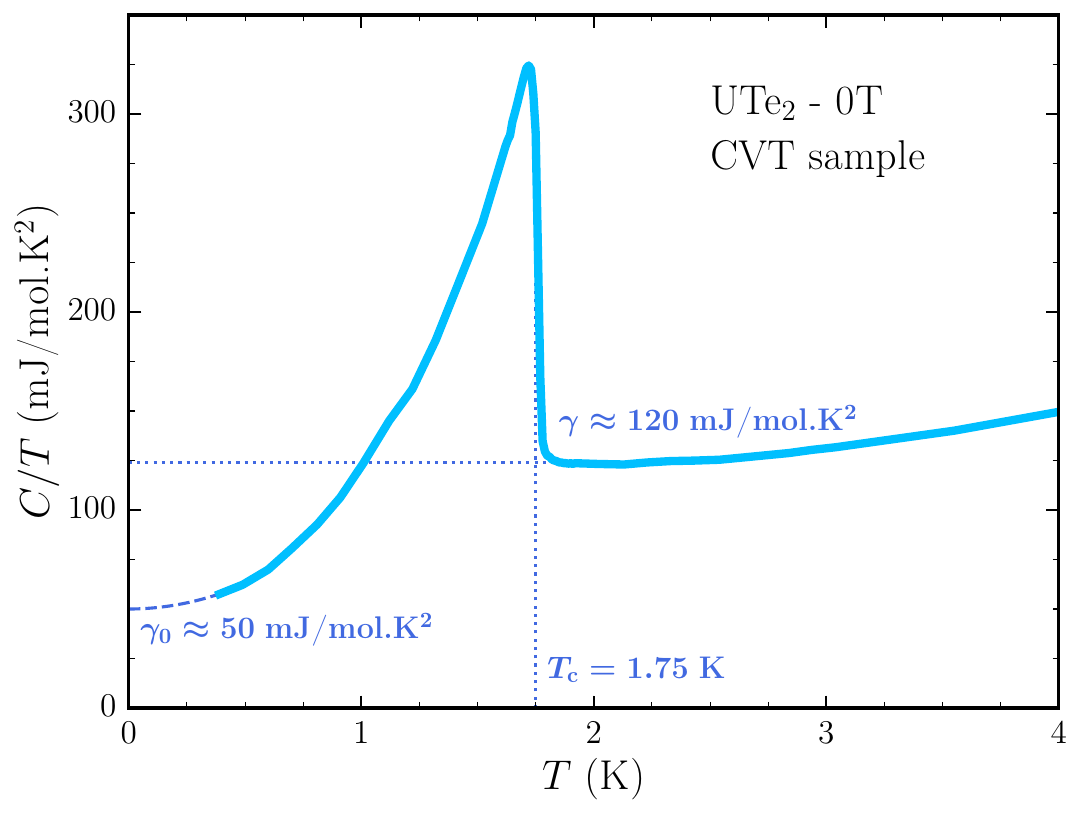}
    \caption{(left) Photograph of the sample before the polishing process. (right) Specific heat divided by temperature as a function of temperature at ambient pressure and zero field.}
    \label{fig:ppms}
\end{figure}

The single crystal used in this study was synthesized using Chemical Vapor Transport in CEA Grenoble. It has been polished by hand to fit inside the pressure cell. The final mass of the sample was $m=23.73~$mg. Figure~\ref{fig:ppms} shows a photograph of the unpolished sample, a specific heat measurement  using a Quantum Design PPMS, and a preliminary analysis, which was then used in the elaboration of the pressure-dependent addenda (see below).

\subsection{Pressure setup}
To measure the absolute values of the heat capacity of UTe$_2$ under hydrostatic pressure, we used a miniature copper-beryllium piston-cylinder pressure cell with an inner diameter of 2~mm. The cell was loaded with the sample, Fluorinert FC-72, as the pressure-transmitting medium, and a piece of lead with a mass of approximately 3~mg, which served as a pressure gauge, inside a Teflon capsule \cite{Nicklas2015}.

We determined the pressure inside the pressure cell by performing a magnetization measurement in a Quantum Design MPMS. For this measurement, we placed a second piece of lead of a similar mass outside the pressure cell. We determined the shift of the superconducting transition temperatures of the lead sample inside the cell with respect to the reference sample outside and converted it to pressure using Ref. \cite{Eiling1981}. Using the differential approach eliminates the influence of remnant magnetic fields from the magnet in the MPMS.

\subsection{Calorimetry setup}
The specific heat was obtained using a compensated quasi-adiabatic heat-pulse technique in a dilution refrigerator. A detailed description of the experimental setup can be found in Refs.~\cite{Nicklas2015} and \cite{Wilhelm2004}. The pressure cell was mounted in a quasi-adiabatic environment, with only a small thermal link to the bath. A heat pulse is applied to the pressure cell, and the resulting temperature rise is recorded. Simultaneously, background heating compensates for the heat leakage to the bath. 

\subsection{Addenda}
\begin{figure}
    \centering
    \includegraphics[width=0.5\linewidth]{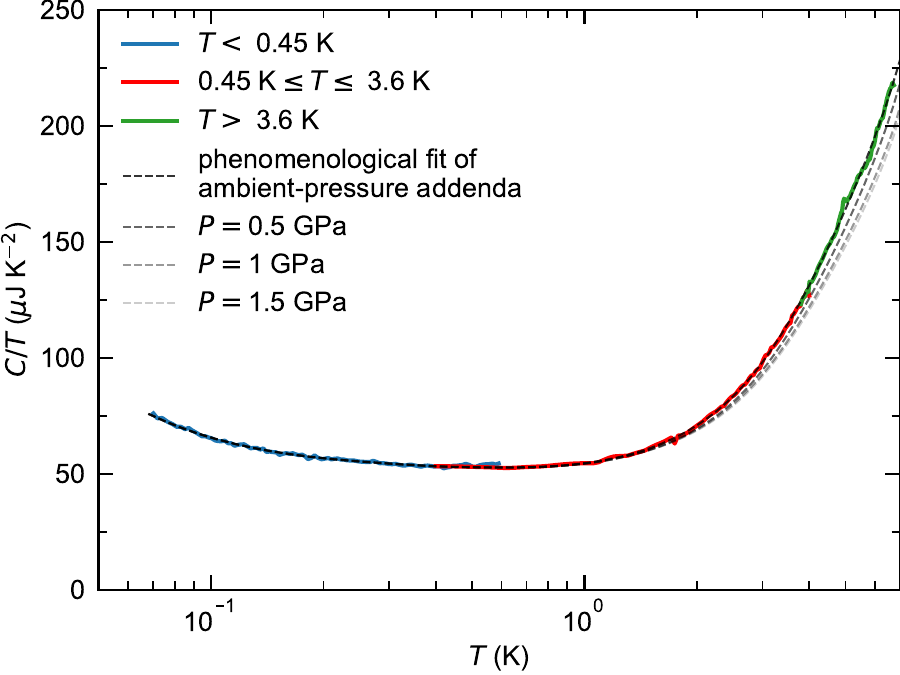}
    \caption{Colored curves present the ambient-pressure addenda constructed from three different approaches  as described in the text for the temperature range $T <\SI{0.45}{\kelvin}$, $\SI{0.45}{\kelvin} \leq T \leq \SI{3.6}{\kelvin}$, and $T>\SI{3.6}{\kelvin}$. Gray dashed lines present the pressure-dependent addenda for three selected pressures as indicated.}
    \label{fig:addenda}
\end{figure}

Accurately determining the addenda is essential to obtaining the absolute heat capacity of the measured sample. Depending on the temperature interval, different methods were used to construct addenda functions for the entire temperature and pressure range of the experiment. The resulting addenda curves from phenomenological fits across the full temperature range are shown in Fig.~\ref{fig:addenda}. To determine the addenda at zero pressure, we divided the experimental temperature range into three intervals. For $T<\SI{0.45}{\kelvin}$, we obtained the addenda directly by measuring the specific heat of the pressure cell, including the Teflon capsule and the lead manometer, but without the UTe$_2$ sample. In the temperature range $\SI{0.45}{K} \leq T \leq \SI{3.6}{\kelvin}$, the addenda was determined by subtracting the specific heat of the sample measured at ambient pressure in the PPMS from the heat-capacity measurement of the pressure cell performed close to ambient pressure. This method is limited to $\SI{3.6}{\kelvin}$, the highest temperature reached in the near-ambient-pressure cell measurement. Above $\SI{3.6}{\kelvin}$, we reconstructed the addenda using the heat capacity measurement with the UTe$_2$ sample at $\SI{1.11}{\giga\pascal}$, taking into account the extra contribution  due to the increase in $\gamma$ deduced from the data for $T<\SI{3.6}{\kelvin}$ and the small pressure dependence of the specific heat of Fluorinert and Teflon \cite{Nicklas_unpublished, Boyer1983}. These contributions were also considered when constructing the addenda functions for the full temperature range at distinct pressures. As depicted in Fig.~\ref{fig:addenda}, they only contribute noticeably to the pressure dependence of the addenda above 1.5 K. Compared to the addenda, the specific heat of the sample in the normal state at 2~K at ambient pressure is only 4\% of that of the addenda, which explains the large scattering of the date in the normal state and to higher temperatures.

\subsection{Low temperature data}

\begin{figure}[ht]
\centering
\includegraphics[width=0.5\linewidth]{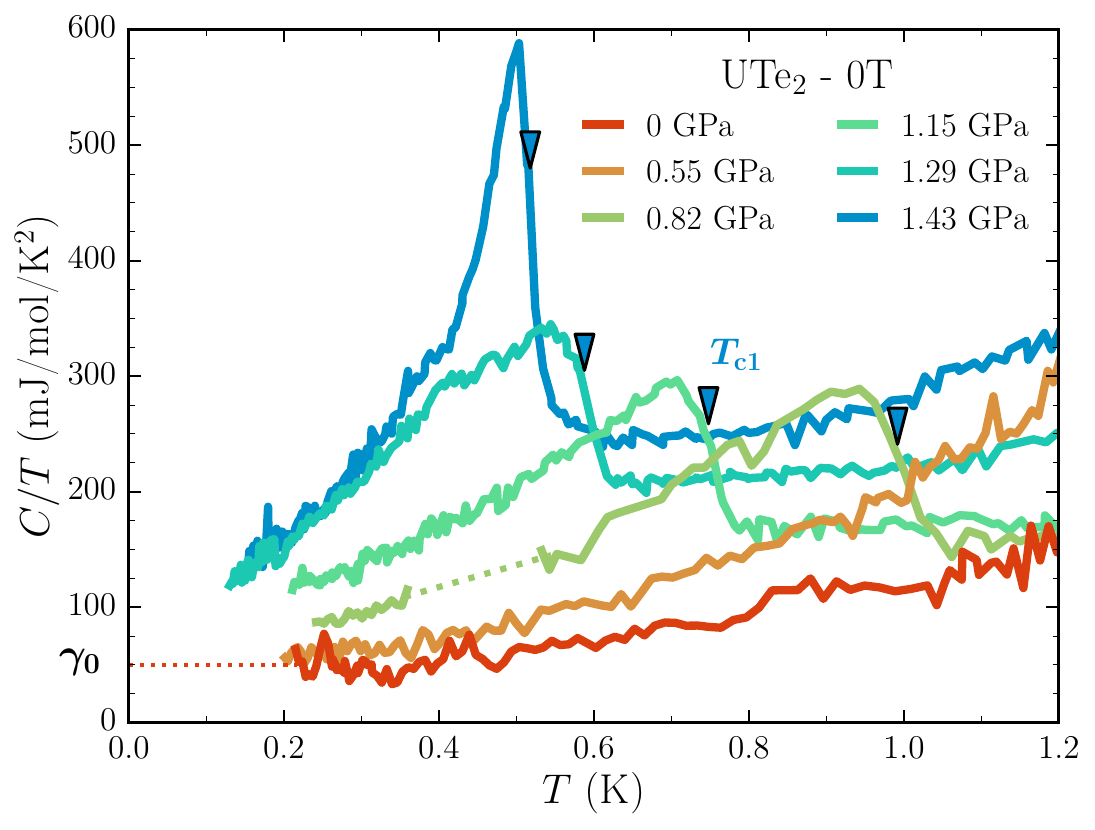}
\caption{\label{fig:zoom}
Specific heat at low temperatures for the pressures up to $\SI{1.4}{\giga\pascal}$.  The blue triangles denote the transition temperature $T_\text{c1}$.}
\end{figure}

Our ambient-pressure data (Fig.~\ref{fig:ppms}) exhibit a sizable residual term $C/T \approx \SI{50}{\milli\joule\per(\mole\cdot\kelvin\squared)}$, as commonly observed in CVT-grown samples. Figure~\ref{fig:zoom} displays the low-temperature specific heat in the superconducting state for all pressures. The extrapolation to $T \rightarrow 0$ has a quite large uncertainty, but seems to yield a residual coefficient $\gamma_0$, that increases  less than  $\gamma$ with pressure. If the ratio $\gamma_0/\gamma$ reflects the fraction of the sample remaining in the normal state, this implies that the superconducting volume fraction of SC1 increases with pressure. This trend suggests that superconductivity becomes progressively less sensitive to sample quality at elevated pressure, consistent with previous resistivity and ac calorimetry studies on samples of varying quality \cite{ajeesh2024}.

\section{3D diagram}
\begin{figure}
    \centering
    \includegraphics[width=0.65\linewidth]{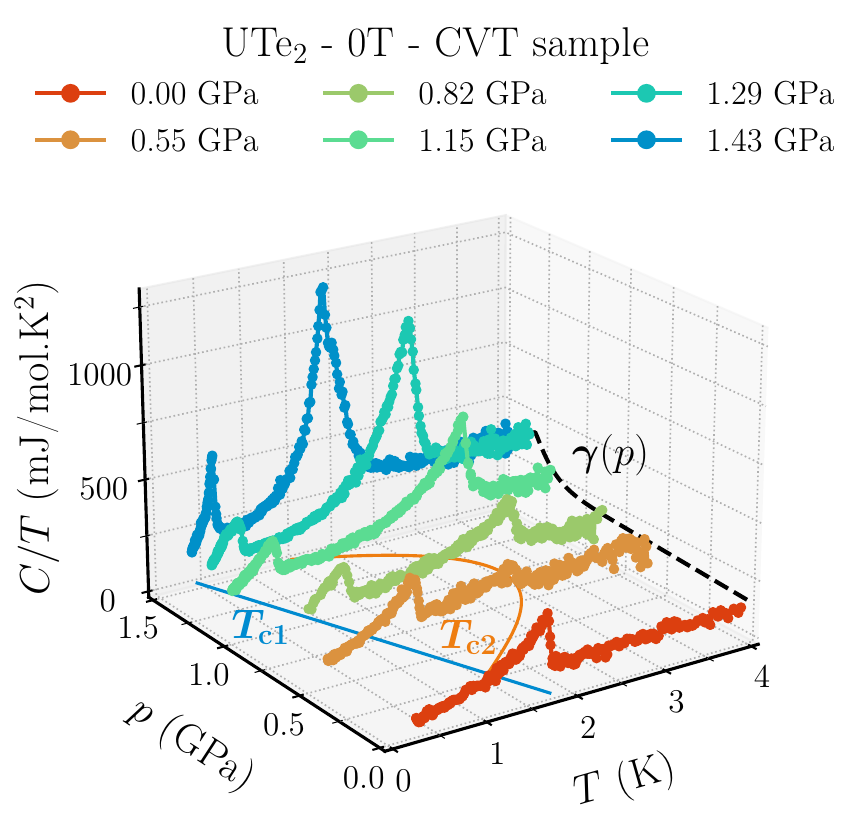}
    \caption{Specific heat below $4~$K and for pressures below the critical pressure. To guide the eye, the trajectories for both $T_\text{c}$, and $\gamma$ are shown respectively as full or dashed lines.}
    \label{fig:3d}
\end{figure}

Figure~\ref{fig:3d} presents the specific data obtained on the CVT sample, showing all superconducting transitions in a single 3D diagram. 

\section{Measurements in diamond anvil cell}
\begin{figure}
    \centering
    \includegraphics[width=0.7\linewidth]{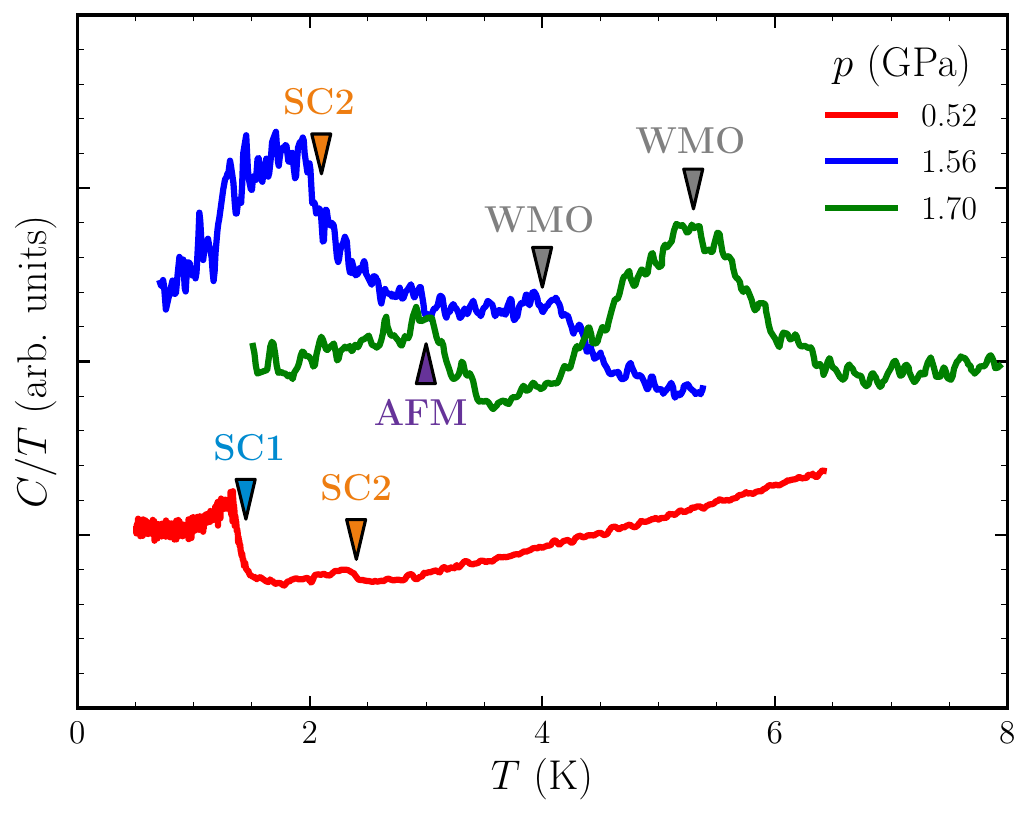} 
    \caption{$C/T$ curves for different pressures obtained in the diamond anvil cell with argon pressure transmitting medium and in-situ pressure tuning.}
    \label{fig:C_DAC}
\end{figure}

A second pressure study of a molten-salt flux (MSF) grown sample was carried out in CEA Grenoble, in a diamond anvil cell with argon as the pressure medium. The pressure was changed in-situ at low temperature and determined  using the ruby fluorescence scale. The sample was heated with an AC current passing through a carbon paste heater on the sample, and the temperature oscillations were measured with a Au/AuFe thermocouple spot-welded on the sample. Figure ~\ref{fig:C_DAC} shows selected curves of $C/T$ obtained. The critical pressure of UTe$_2$ is systematically higher when measured in a diamond anvil cell with argon compared to piston cylinder cells with Daphne oil or Fluorinert used here, so to compare the two experiments (Fig.~2 in main text) the pressure scales were normalized by the experimental critical pressures $p_\textsc{c}$. 

\section{Details on data analysis}

\subsection{Superconducting transitions}
To determine the transition temperatures $T_\textsc{c1}$ and $T_\textsc{c2}$  for the superconducting phases, we used a Gaussian-fitting model, following the same procedure as in \cite{vasina_connecting_2025}. This is also useful to follow the characteristics of the specific heat anomalies as a function of pressure, particularly the jump height $\Delta C/T$ in a consistent way. For the curve at $\SI{1.43}{\giga\pascal}$  where the WMO phase as well as SC1 and SC2 are present, the shape of the transitions did not allow us to fit a mean-field, second-order transition, so both the $T_\text{c}$ and $\Delta C$ are taken graphically. 

\subsection{Magnetic transitions}
Both, AFM and WMO transition temperatures were taken at the maximum of the corresponding specific heat anomalies.

\subsection{Determination of the \texorpdfstring{$\gamma$}{gamma} coefficient}

For pressures up to $\SI{1.29}{\giga\pascal}$, the Sommerfeld coefficient $\gamma$ is determined from the fit of the high temperature superconducting transition above the critical temperature. For the two highest pressures, $\gamma$ is estimated graphically from the $C/T$ data in the temperature range above the WMO transition.

A phonon contribution $\beta T^2$ was subtracted from $C/T$, with $\beta = 2.3~$mJ/(mol.K$^{4}$) determined from the ambient-pressure PPMS measurement and assumed to remain essentially unchanged over the pressure range explored here.

\subsection{Empirical model for the pressure evolution of the specific heat coefficient}
\begin{figure}
    \centering
    \includegraphics[height=0.31\linewidth]{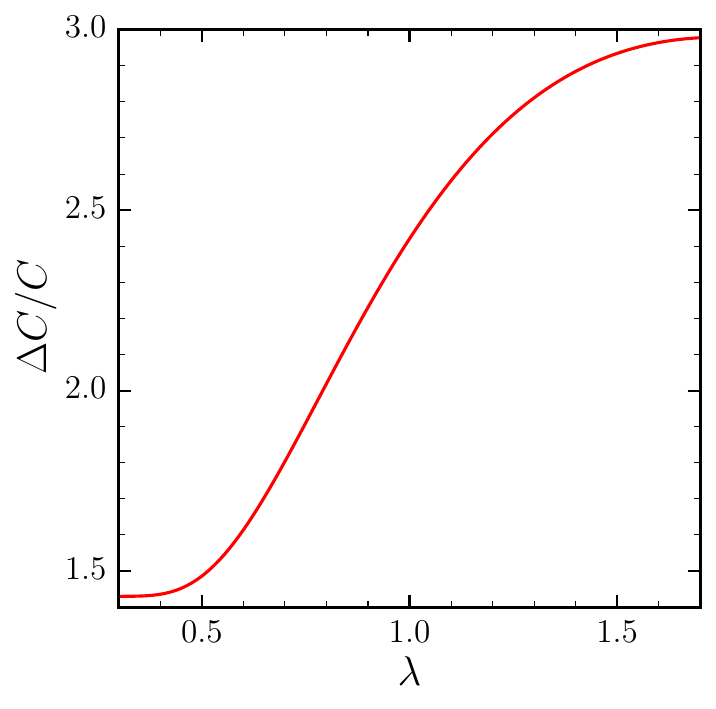}\quad
    \includegraphics[height=0.31\linewidth]{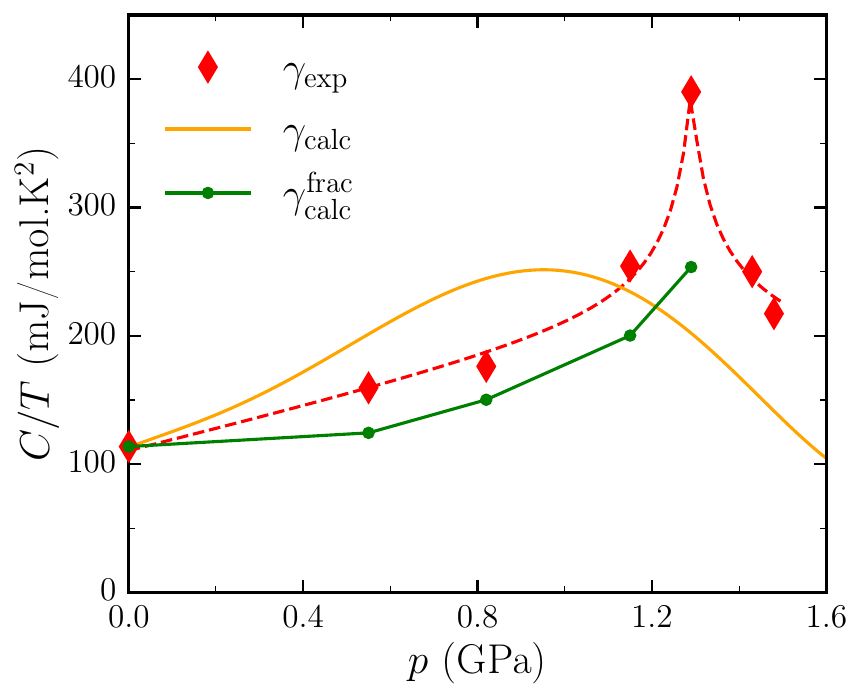}\quad
    \includegraphics[height=0.31\linewidth]{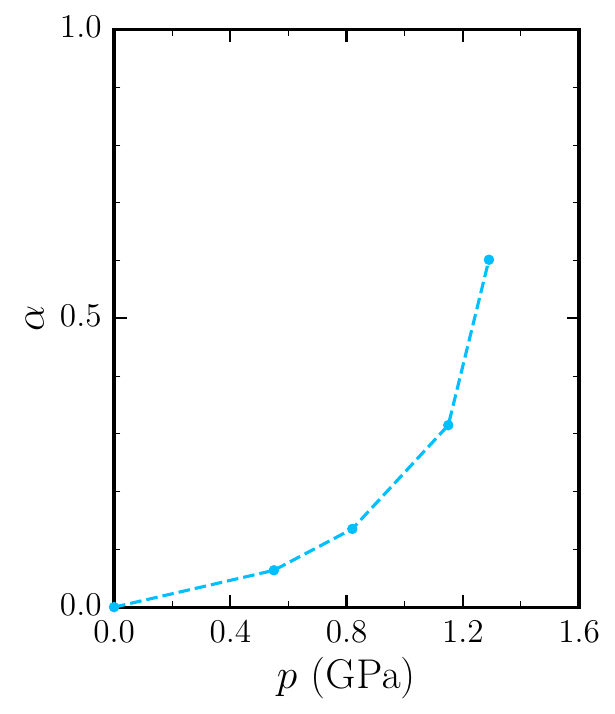}
    \caption{(Left)Relative specific heat jump as a function of the strong coupling parameter $\lambda$ calculated with the empirical formula from \cite{MarsiglioPRB1986} for $\mu^{\ast}$=0.1. (Middle) Comparison between the experimental values of $\gamma(p)$   and predictions from a renormalization by the pairing mechanisms deduced from the pressure dependence of the critical temperatures of SC1 and SC2 ($\gamma_{\text{calc}}$: plain orange line). This is compared to the calculation taking into account the fraction $\alpha$ of the Fermi surface on which SC2 nucleates  ($\gamma_{\text{calc}}^{frac}$: green line). Dotted red line is a guide to the eye. (Right) pressure dependence of $\alpha$ determined using the SC2 specific heat jump.}
    \label{fig:Marsiglio}
\end{figure}

As explained in the main text, we expect that the pressure evolution of the $\gamma$ coefficient, the critical temperatures, the specific heat jumps, are related, being all controlled by the evolution of the pairing mechanisms.
Indeed, from Eliashberg theory, we know that the superconducting critical temperature $T_\text{c}$ depends on a typical frequency $\Omega$ of the fluctuations responsible for the pairing, of a Coulomb repulsive pseudo potential $\mu^{\ast}$ usually in the range $0.1-0.15$, and a strong coupling constant $\lambda$ arising from the density of interactions and the fluctuation spectrum, averaged over the Fermi surface. 
Unlike BCS theory, there is no general analytical formula expressing $T_\text{c}$ as a function of these parameters. As a guide for the dependence on these parameters. we can use the empirical McMillan formula:
\begin{equation}
    T_\text{c} = \Omega \exp \left( - \frac{ 1 + \lambda}{\lambda - \mu^{\ast}} \right)
    \label{equ:McMillan}
\end{equation}
The Eliashberg theory, well established for the electron-phonon pairing mechanism, allows the determination of $T_\text{c}$ from a microscopic description of the phonon spectrum and of the electron-phonon interaction. 
It also predicts the renormalization of the normal state properties, and notably an enhancement of the specific heat coefficient by a factor (1+$\lambda$). 
Such a renormalization is generally supposed to hold also for magnetic pairing mechanisms. 
Moreover, the height of the specific heat jump grows also beyond the BCS value in the strong  coupling regime (when $\lambda$ is much larger than $\mu^{\ast}$). 
There is again no general analytical formula for this jump, but like for the Mc-Millan formula 
(Eqn.~\ref{equ:McMillan}), there are empirical expressions for this jump as a function of $T_\text{c}/\Omega$ \cite{MarsiglioPRB1986}. 
Figure~\ref{fig:Marsiglio} shows how the relative jump depends on the strong coupling constant $\lambda$ for a fixed value of $\mu^{\ast} = 0.1$.
Let us call $\Psi(\lambda, \mu^\ast)$ this function giving $\Delta C/C(\lambda, \mu^\ast)$ for a fully gapped state (only conventional superconductors were considered in Ref.~\cite{MarsiglioPRB1986}). As shown in Fig.~\ref{fig:Marsiglio} (left), it seems to saturate for values of $\lambda > 1.6$.

We used this dependence ($\Psi(\lambda,\mu^\ast)$) of the relative specific heat jump to fix a value of  $\lambda_{\text{SC1}}(p=0)$ appropriate for the SC1 phase: 
we considered that $\Delta C/C$ is of order 2 for the best molten flux samples, yielding $\lambda \approx 0.7-0.8$, and that the reduction of $\Delta C/C$ in other samples (like the present one) arise from pair-breaking effects, 
also generating the residual terms for $C/T$ at the lowest temperatures (see Fig.~\ref{fig:zoom}). 
We then used the same strong coupling model as in \cite{Rosuel2023} to get the pressure dependence $\lambda_{\text{SC1}}(p)$ in order to reproduce the decrease of $T_\text{c}$ for the SC1 phase.
In principle, this is not appropriate as the Eliashberg model used here is valid for a transition from the normal state to the superconducting state, not for a transition between two superconducting states.
Hence, at pressures $p \geq \SI{0.55}{\giga\pascal}$ the model should not apply. 
However, because the slope $\frac{dT_\text{c}}{dp}$ does not change when the SC1 and SC2 phase lines cross, we considered that this should be "good enough" at the empirical level we are considering here.

Regarding the SC2 phase, we know that if its pairing mechanism is responsible for the increase of $\gamma$ under pressure (despite the small decrease of $\lambda_{\text{SC1}}(p)$),  $\lambda_{\text{SC2}}$ needs to be rather large under pressure so that the factor $(1+\lambda_{\text{SC2}}$) renormalizing the bare specific heat coefficient reaches values of order 3. 
Moreover, a value of $\lambda_{\text{SC2}}$ of order 3 is probably among the largest ever observed. 
And as opposed to the SC1 phase, because the maximum or $T_\text{c2}$ seems little sample dependent (see e.g. Fig.~2 of the main text), we introduced no pair-breaking parameter in the calculation of $T_\text{c2}(\lambda)$.
Then, targeting a value $\lambda_{\text{SC2}} \approx 3$ for the pressure maximum of $T_\text{c2}$ 
yields the curve shown on Fig.~3(c) of the main text. 
Extrapolation of $\lambda_{\text{SC2}}$ toward zero pressure is used only to draw the pressure evolution of $\gamma$ induced by the renormalization from both pairing mechanisms of the SC1 and SC2 phases through:
\begin{equation}
    \gamma_{\text{calc}} = \gamma_0 (1 + \lambda_{\text{SC1}})(1+\lambda_{\text{SC2}})
    \label{equ:gammaSC}
\end{equation}
Here, $\gamma_0$ would be a bare Sommerfeld coefficient, without the mass renormalization induced by the pairing mechanisms of SC1 and SC2.
Fig.~\ref{fig:Marsiglio}(middle) shows how badly a $\gamma$ term calculated in such way (orange line) matches experimental data (red diamonds). 
Changing the starting values of $\lambda_{\text{SC1}}$ or $\lambda_{\text{SC2}}$ would in no way change the predicted decrease of $\gamma_{\text{calc}}$ past the maximum of $T_\text{c}$ or the smooth maximum  of $\gamma_{\text{calc}}$: both are imposed by the relation (see Eqn.~\ref{equ:McMillan}) between $T_\text{c}$ and $\lambda$.

In order to take account of the vanishing specific heat jump of the SC2 phase at low pressures, we can add the hypothesis that the SC2 phase emerges first on a very small portion of the Fermi surface. This can be seen as an extreme case of multiband superconductivity, with negligible interband interactions. 
By contrast, the pairing mechanism of SC1 develops (as usual) on the whole Fermi surface, as demonstrated by the vanishing residual term of the zero field specific heat on the best UTe$_2$ crystals \cite{aoki_molten_2024}. 
Let us call "$\alpha$" ($0 \leq \alpha \leq 1$) the fraction of the Fermi surface (density of states) affected by the pairing mechanism of SC2.
Then, mass renormalization by this mechanism is effective only on a fraction $\alpha$ of the electronic specific heat coefficient.
The pressure dependence of $\gamma$ would then not follow Eqn.~\ref{equ:gammaSC} but instead: 
\begin{equation}
    \begin{aligned}
    \gamma &= (1 - \alpha) \gamma_0 (1 + \lambda_{\text{SC1}}) + \alpha \gamma_0 (1 + \lambda_{\text{SC1}})(1+\lambda_{\text{SC2}}) \\
    &=   \gamma_0 (1 + \lambda_{\text{SC1}})( 1 + \alpha \lambda_{\text{SC2}})  
     \end{aligned}
    \label{equ:gammaAlpha}
\end{equation}
In the same spirit, the specific heat jump of the SC2 phase should happen only relative to the fraction $\alpha$ of $\gamma$. 
Hence it means that we can take the specific heat jump as a measure of $\alpha$ at the different pressures. 
At this empirical level, everything is determined once $\lambda_{\text{SC2}}(p)$ and $\lambda_{\text{SC1}}(p)$ are known (from the respective $T_\text{c}(p)$):
\begin{equation}
    \begin{aligned}
    \Delta C(p) &= \Psi(\lambda_{\text{SC2}}(p),\mu^{\ast}) \times  \alpha\gamma_0[1+\lambda_{\text{SC1}}(p)][1+\lambda_{\text{SC2}}(p)] \\
\alpha &= \frac{\Delta C(p)}{\gamma_0[1+\lambda_{\text{SC1}}(p)][1+\lambda_{\text{SC2}}(p)]} \frac{1}{\Psi(\lambda_{\text{SC2}}(p),\mu^{\ast})}\\
\gamma_{\text{calc}} &= \gamma_0[1+\lambda_{\text{SC1}}(p)] + \frac{\Delta C(p)}{\Psi(\lambda_{\text{SC2}}(p),\mu^{\ast})} \frac{\lambda_{\text{SC2}}(p)}{1+\lambda_{\text{SC2}}(p)} 
     \end{aligned}
    \label{equ:DeltaCAlpha}
\end{equation}
And owing to the rapid saturation of $\psi(\lambda,\mu^\ast)$ for $\lambda \geq 1.6$, and the slow variation of $\lambda_{\text{SC1}}$, the new expression in Eqn.~\ref{equ:DeltaCAlpha} shows that in the strong coupling regime, $\gamma_{\text{calc}}$ is essentially determined by $\Delta C$, with a weak dependence on the exact values of $\lambda_{\text{SC2}}(p)$.
Expression Eqn.~\ref{equ:DeltaCAlpha} was used to draw $\gamma_{\text{calc}}$ on the Fig.~3(a) of the main text, assuming that $\alpha = 0$ at zero pressure so that (see Equ.~\ref{equ:gammaAlpha}) $\gamma_0 = \gamma_{\text{exp}}(p=0)/(1 + \lambda_{\text{SC1}}(p=0))$. This updated value for $\gamma_{\text{calc}}$ is also shown on Fig.~\ref{fig:Marsiglio}(middle), labeled as $\gamma_\text{calc}^\text{frac}$. The values for $\alpha$ deduced from the specific heat jump data using Eqn.~\ref{equ:DeltaCAlpha} are plotted in Fig.~\ref{fig:Marsiglio}(right).

By contrast, due to the direct dependence of $\gamma_{\text{calc}}$ on the specific heat jump, we did not try to extend these "predictions" at pressures where the SC2 phase emerges from a normal state with WMO. Indeed, we discuss in the main text that even the shape of the transition strongly changes in this pressure region, suggesting that both superconductivity and magnetism are then contributing to $\Delta C$. In such a case, the "strong coupling estimate" $\frac{\Delta C}{C} = \Psi(\lambda,\mu^\ast)$ should break down.

In principle, an even better agreement between this model and the experiments could be expected if the SC2 phase has nodes, so that  $\frac{\Delta C}{C}$ is overestimated by $\Psi(\lambda, \mu^\ast)$.
However, a more interesting issue is to understand if the picture given by this simple model, which seems rather successful to link our absolute measurements of the specific heat jump and of the $\gamma$ value under pressure, has a real physical meaning. 
Indeed, the main problem remains to understand how superconductivity can arise with such a small specific heat anomaly: upper critical field measurements of the SC2 phase \cite{Braithwaite2019, Knebel2020} clearly show that it is also governed by heavy quasiparticles, so that the initial smallness of the specific heat jump cannot be attributed to its development on "light bands".
Then the difficulty is to understand how a large value of $\lambda_{\text{SC2}}$ can emerge already when the SC1 and SC2 line cross, if only a very small fraction of the Fermi surface is involved in the pairing mechanism.
As mentioned in the main text, this is similar to the situation encountered in low carrier superconducting systems like SrTiO$_3$\cite{CollignonARCMP2019} or semi-metals like YPtBi \cite{KimSciAdv2018}.
The issue is not always clearly solved for these systems, but in the case of SrTiO$_3$, it has been shown early on that a very strong electron phonon pairing involving plasmon modes might compensate the small density of states \cite{TakadaJPSJ1980}.
Naturally, such a mechanism is not relevant for UTe$_2$.
Presently, a microscopic picture explaining the behavior of $\Delta C/C$ in the SC2 phase, well confirmed by the present absolute specific heat measurements, is still missing. 
\bibliography{UTe2.bib}	